\documentclass[a4paper,11pt]{article}
\usepackage{geometry}
\usepackage{a4wide}
\usepackage{graphicx}
\usepackage{epsf}
\usepackage{amsmath}
\usepackage{amssymb}
\usepackage[normalem]{ulem}

\usepackage{cite}
\usepackage{multirow}
\usepackage{appendix}
\usepackage{tikz}
\usepackage{amsfonts,amsthm,euscript,braket,xcolor}
\newcommand{\be}{\begin{equation}}
\newcommand{\ee}{\end{equation}}

\newcommand{\Rmnum}[1]{\expandafter\@slowromancap\romannumeral #1@}
\newcommand{\bea}{\begin{eqnarray}}
\newcommand{\eea}{\end{eqnarray}}

\numberwithin{equation}{section}

\usepackage{ulem}

\begin{document}


\title{\bf Entropic force and real-time dynamics of holographic quarkonium in a magnetic field}

\author{ \textbf{\textbf{Siddhi Swarupa Jena$^{a}$}\thanks{519ph2015@nitrkl.ac.in}, \textbf{Bhaskar Shukla$^{a}$}\thanks{519ph1003@nitrkl.ac.in}, David Dudal$^{b,c}$}\thanks{david.dudal@kuleuven.be},
, \textbf{Subhash Mahapatra$^{a}$}\thanks{mahapatrasub@nitrkl.ac.in},
 \\\\\textit{{\small $^a$ Department of Physics and Astronomy, National Institute of Technology Rourkela, Rourkela - 769008, India}}\\
\textit{{\small $^b$ KU Leuven Campus Kortrijk--Kulak, Department of Physics, Etienne Sabbelaan 53 bus 7657,}}\\
\textit{{\small 8500 Kortrijk, Belgium}}\\
\textit{{\small $^c$  Ghent University, Department of Physics and Astronomy, Krijgslaan 281-S9, 9000 Gent, Belgium}}
}

\date{}

\maketitle
\abstract{ We continue the study of a recently constructed holographic QCD model supplemented with magnetic field. We consider the holographic dual of a quark, anti-quark pair and investigate its entropy beyond the deconfinement phase transition in terms of interquark distance, temperature and magnetic field. We obtain a clear magnetic field dependence in the strongly decreasing entropy near deconfinement and in the entropy variation for growing interquark separation. We uncover various supporting evidences for inverse magnetic catalysis. The emergent entropic force is found to become stronger with magnetic field, promoting the quarkonium dissociation. We also probe the dynamical dissociation of the quarkonium state, finding a faster dissociation with magnetic field.  At least the static predictions should become amenable to a qualitative comparison with future lattice data. }

\section{Introduction}
\label{sec1}
The dissociation of heavy quarkonium bound states, such as charmonium $c\bar{c}$ and bottomonium $b\bar{b}$, has long been suggested as an indication of the formation of a hot and deconfined quark-gluon plasma (QGP). It is believed that the heavy quarkonium bound states are an important probe of the finite temperature QCD matter and might serve as a signature of deconfinement \cite{Matsui:1986dk}. In the experimental facilities, not only a strong suppression of heavy quarkonium near the deconfinement transition has been found at recent Relativistic Heavy Ion Collider (RHIC) and the Large Hadron Collider (LHC) but also a stronger suppression at lower  than at larger energy density has been observed \cite{PHENIX:2006gsi,ALICE:2013osk}. The strong belief that these observations might not only shed new light on disclosing the QCD behaviour near the deconfinement temperature but also might provide accurate information about its critical points and phase diagram, has led to an intense investigation into the nature of the quarkonium suppression from various perspectives, be it from theory, lattice or experiment sides.
\begin{figure}[t!]
\begin{minipage}[b]{0.5\linewidth}
\centering
\includegraphics[width=2.8in,height=2.3in]{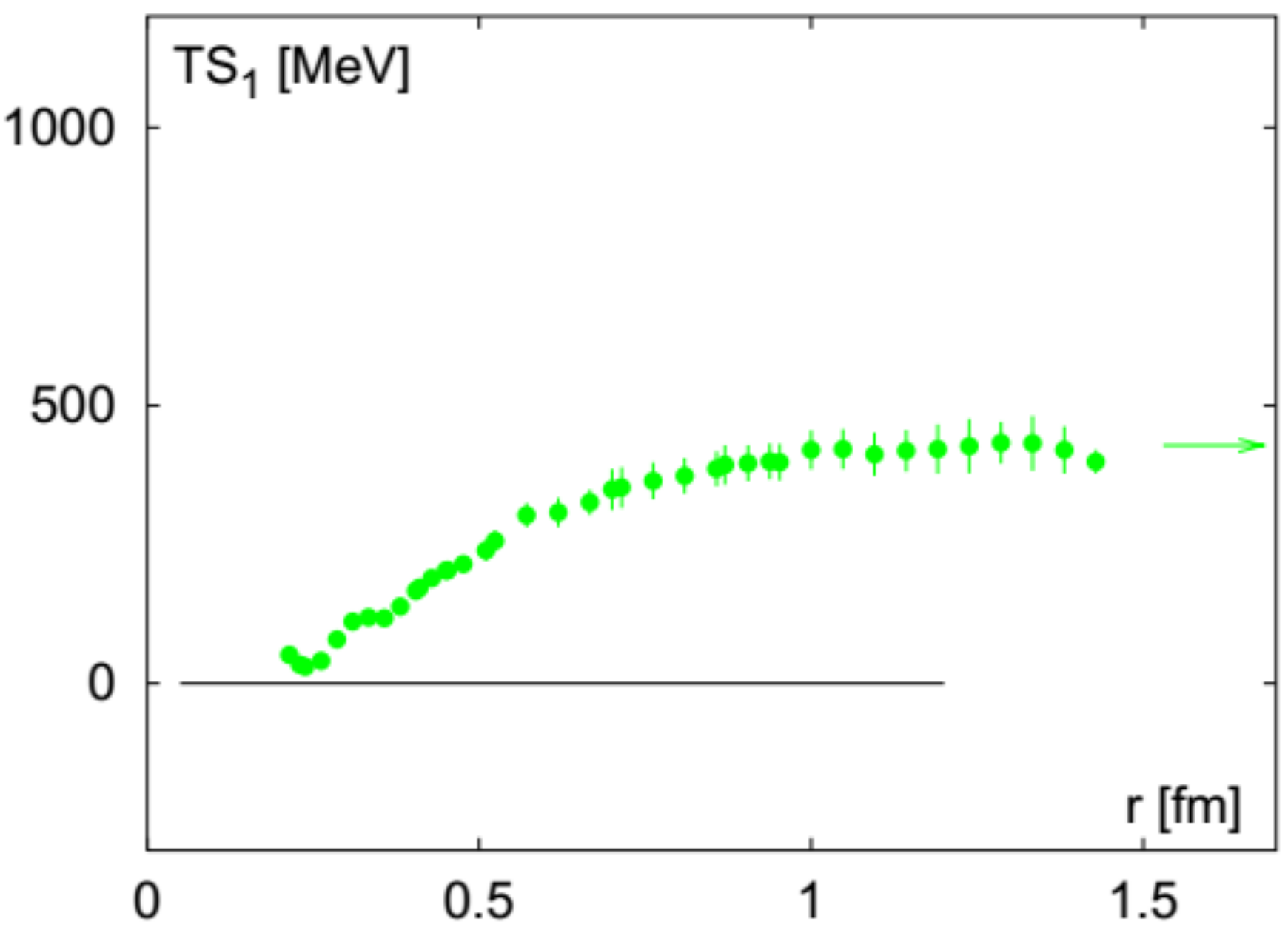}
\caption{ \small Two flavour lattice QCD result for the entropy of a $q\overline q$ pair as function of quark-antiquark separation at temperature $T\simeq 1.3T_c$. The figure is taken from \cite{Kaczmarek:2005zp}.}
\label{latticeQCD2}
\end{minipage}
\hspace{0.4cm}
\begin{minipage}[b]{0.5\linewidth}
\centering
\includegraphics[width=2.8in,height=2.5in]{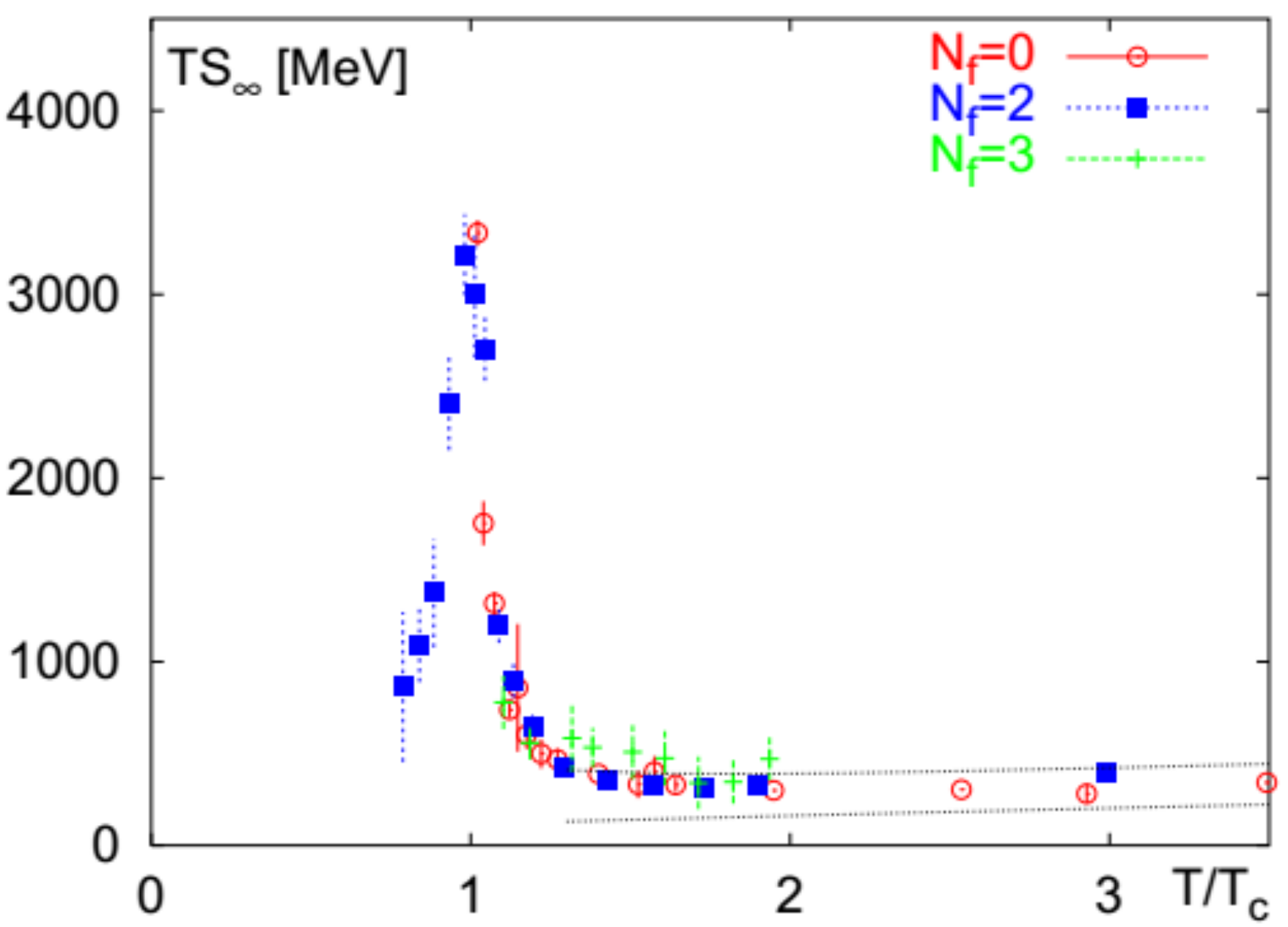}
\caption{\small Lattice QCD result for the entropy of a $q\overline q$ pair as function of temperature $T/T_c$ for large $q\overline q$ separation. The figure is taken from \cite{Kaczmarek:2005zp}. }
\label{latticeQCD1}
\end{minipage}
\end{figure}

Various theoretical scenarios such as the colour screening mechanism \cite{Matsui:1986dk}, a recombination of the produced heavy quarks \cite{Braun-Munzinger:2000csl,Thews:2000rj}, the imaginary potential mechanism \cite{Laine:2006ns}, etc., have been proposed for its explanation. One such scenario for the quarkonium suppression, based on earlier lattice QCD results of the quark-antiquark $q\bar{q}$ pair entropy $S$, was proposed in \cite{Kharzeev:2014pha}. The relevant lattice QCD results are summarized in Figures~\ref{latticeQCD2} and \ref{latticeQCD1} \cite{Kaczmarek:2002mc,Petreczky:2004pz,Kaczmarek:2005zp}. The important observations are i) there appears a large and a sharp peak in the entropy of the heavy $q\bar{q}$ pair at the deconfinement temperature, ii) the entropy decreases very fast with temperature $T$ away from the deconfinement temperature, and iii) the entropy first gradually increases with the interquark distance $\ell$ and then saturates to a constant value at large separations. Importantly, this high entropy of the $q\bar{q}$ pair cannot arise entirely from the hot QCD plasma and should come from additional (extended)
gauge field configurations \cite{Hashimoto:2014fha}. Based on these observational facts, an ``emergent entropic force''
\begin{eqnarray}
F = T \frac{\partial S}{\partial \ell}  \,,
\label{entropicforce}
\end{eqnarray}
mechanism was proposed to explain the quarkonium suppression \cite{Kharzeev:2014pha}. In particular, the fact that $S$ increases with $\ell$ can provide a positive repulsive entropic force and can boost the self-destruction of the quarkonium bound state, thereby causing suppression. Moreover, the dissociation of a bound state to a delocalized state was shown to be maximal near the deconfinement temperature,  thereby providing a potential explanation for the strong suppression of heavy quarkonium around that temperature.

Another interesting QCD quantity, which is directly related to the nature of quarkonium suppression, is the dissociation time of the quark-antiquark pair. It provides an important time scale in the physics of quarkonium suppression and is expected to play an important role in establishing the correct mechanism of QGP formation.  However, this quantity is inherently related to the real-time dynamics of the quarkonium, and therefore, is very hard to compute even in lattice settings due to its inherently Euclidean (imaginary time) formulation.

In recent years the investigation of the effects of an electromagnetic field, in particular a magnetic field, on QCD related phenomena has attracted a lot of attention. It has recently been suggested that a very strong magnetic field might be produced in the early stages of noncentral relativistic heavy-ion collisions and can have important consequences for QCD phases \cite{Skokov:2009qp,Bzdak:2011yy,Voronyuk:2011jd,Deng:2012pc,DElia:2010abb,DElia:2021tfb,Tuchin:2013ie}. The maximum strength of the produced magnetic field is estimated to be around $eB\backsim 0.3~\text{GeV}^2$. Though this magnetic field decays after the collision, however, importantly, it is expected to remain sufficiently high when the QGP forms, thence can leave its imprint on various QGP observables \cite{Tuchin:2013apa,McLerran:2013hla}. In fact, a large magnetic field (though less intense) of order $eB\backsim 1~\text{MeV}^2$ is also believed to exist inside the neutron stars \cite{Duncan:1992hi}, whereas a much stronger field of order $eB\backsim 4~\text{GeV}^2$ is expected to have been generated in the early Universe \cite{Vachaspati:1991nm}. Therefore, unsurprisingly, thanks to a large range of physical situations where the magnetic field can play a significant role, a lot of research on related subjects, ranging from cosmology to early Universe physics, have been carried out in the last few years; see \cite{Miransky:2015ava,Kharzeev:2012ph,Ferrer:2012wa,Gatto:2012sp,Iwasaki:2021nrz} for detailed reviews on these subjects.

In the context of QCD, the magnetic field has many observable consequences. It has been shown to play a destructive role in the deconfinement and chiral transition temperatures (aka the inverse magnetic catalysis) \cite{Bali:2011qj,Bali:2012zg,Ilgenfritz:2013ara,Bruckmann:2013oba,Fukushima:2012kc,Ferreira:2014kpa,Mueller:2015fka,Bali:2013esa,Fraga:2012fs,Ayala:2014iba,Ayala:2014gwa,Fraga:2012ev}. It also has direct consequences on the string tension of heavy quarks, particularly, causing suppression/enhancement of the string tension in a direction parallel/transverse to the magnetic field \cite{Bonati:2014ksa,Bonati:2016kxj,DElia:2021tfb,Simonov:2015yka}. It is also expected to influence the charge dynamics in QCD, thereby providing anomaly induced novel dynamical phenomena such as the chiral magnetic and chiral vortical effects \cite{Fukushima:2008xe,Kharzeev:2007jp,Kharzeev:2015znc,Kharzeev:2010gd,STAR:2021mii}, which in turn may have important consequences on the strong CP problem and baryon asymmetry generation in the early Universe \cite{Kharzeev:2020jxw}.  It can also induce more thermodynamic phases and ordinary types of charge transports in QGP, and is also essential for the anisotropic hydrodynamic description of QCD matter \cite{Das:2016cwd,Chatterjee:2018lsx,Gursoy:2014aka,Giataganas:2017koz}.

It is clear that the magnetic field has emerged as a major player in QGP and deconfinement related physics. It is, therefore, crucial to investigate how this magnetic field affects the quarkonium dissociation and its related time scale, particularly, from the point of view of the entropic force. For other works on charmonium properties in a magnetic field, see \cite{Marasinghe:2011bt,Dudal:2014jfa,Cho:2014loa,Guo:2015nsa,Yoshida:2016xgm,Amal:2018qln,Zollner:2020nnt,Iwasaki:2021nrz}.

Indeed, lattice supported results of the non-trivial dependence of the quark-antiquark free energy and string tension on the magnetic field do indicate substantial changes in the structure of the quark-antiquark entropy and the quarkonium dissociation time at finite magnetic field compared to its zero counterparts. Moreover, one could naturally also expect substantial changes in the dissociation structure depending on whether the pair is oriented parallel or perpendicular to the magnetic field.

Unfortunately, the near deconfined region of QCD is dominated by non-perturbative physics and remains challenging to probe. The entropy near the deconfinement temperature measured from a mean-field Debye screening approximation is found to be an order of magnitude smaller than that of lattice results, implying the breakdown of the weak coupling (perturbative) approximation \cite{Hashimoto:2014fha}. The numerical oriented approach of lattice QCD can be a compelling tool here. However, not only is it difficult to perform a real-time analysis of the dissociation via lattice simulations (due to its inherent Euclidean nature), but it is also too expensive for the entropy calculation (as it requires an entire thermal scan of the free energy). Another tool that is suitable for computations at strong coupling is the gauge/gravity duality \cite{Maldacena:1997re,Gubser:1998bc,Witten:1998qj}. The idea that certain quantum field theories at strong couplings are dual to the classical gravitation theories in AdS spaces in one more dimension has provided an elegant technique to probe QCD related physics. Indeed, one of the original aims of this duality is to probe non-perturbative, strongly coupled gauge theories, and undoubtedly by now, many essential features of QCD have been reproduced from it, and in some situations, new and striking insights into the domain of strongly coupled QCD have also been made, for some interesting reviews let us refer to \cite{Casalderrey-Solana:2011dxg,Gubser:2009md,Jarvinen:2021jbd,Gursoy:2010fj}.

Accordingly, the motivation of this work is to explore the anisotropic effects of a magnetic field on the quark-antiquark free energy, entropy and dissociation time using AdS/QCD holography. For this purpose, we concentrate on a particular bottom-up magnetised AdS/QCD model of \cite{Bohra:2019ebj,Bohra:2020qom,Dudal:2021jav}.  This model is based on Einstein-Maxwell-dilaton (EMD) gravity and its novelty lies in the fact that not only the gravity equations can be solved analytically but also its dual boundary theory exhibits several lattice supported magnetised QCD features. A short survey of this model, along with its relevant QCD features, is presented in the next section for convenience \footnote{One can also consider the top-down models and, in particular, consider the simplest AdS magnetised background of \cite{DHoker:2009mmn,DHoker:2009ixq}, which was constructed from Einstein-Maxwell gravity. This model, however, does not exhibit a Hawking/Page type phase transition between the black hole and thermal-AdS, and hence, there is no confinement/deconfinement phase transition in the dual boundary theory. This problem can however be bypassed by introducing an additional dilaton field, without having a kinetic term, in the action, i.e., using a soft wall type model phenomenology \cite{Dudal:2015wfn}. Unfortunately, these models are not entirely satisfactory in a way as they do not solve all the gravity field equations consistently.}. We find that the $q\bar{q}$ free energy exhibits a Coulombic structure at small interquark separations whereas it flattens out at large separations for all temperatures and magnetic fields. The size of the $q\bar{q}$ bound state, corresponding to the maximally allowed (connected) string length, at the deconfinement temperatures is found to be decreasing with the magnetic field. This result is true irrespective of the orientation of the magnetic field. We further analyse the entropy of the $q\bar{q}$ pair and find that it not only exhibits a sharp rise and a peak near the deconfinement temperature but also saturates to a constant value at large interquark separations for all magnetic field values. For vanishing magnetic field, these findings are in qualitative agreement with known lattice QCD results. With magnetic field, the magnitude of the $q\bar{q}$ entropy decreases substantially near the deconfinement temperature compared to the zero field case. Similarly, its saturation value at large separation decreases with magnetic field. Interestingly, the strength of the entropic force increases with magnetic field near the deconfinement temperature, suggesting an enhanced dissociation of heavy quarks in the presence of a magnetic field at that temperature. These results point to the inverse magnetic catalysis response of the QCD system to a magnetic field \cite{Bali:2011qj}. We further analyse the real-time dynamics of quarkonium dissociation and calculate the dissociation time for different temperatures and magnetic fields. We find that the dissociation time gets smaller for higher temperatures and magnetic fields. In particular, the magnetic field enhances the dissociation rate. Again, the holographic QCD system exhibited these features for both parallel and perpendicular orientations of the magnetic field.

Before we focus on this specific magnetised AdS/QCD model, let us also  mention here other top-down and bottom-up models where further discussion on the interplay between the magnetic field and QCD observables have been explored  \cite{Albash:2007bk,Johnson:2008vna,Callebaut:2013ria,Dudal:2016joz,Preis:2010cq,Filev:2010pm,Dudal:2015wfn,Mamo:2015dea,Li:2016gfn,Evans:2016jzo,Bolognesi:2011un,Ballon-Bayona:2017dvv,Rodrigues:2017cha, Rodrigues:2017iqi,McInnes:2015kec,Dudal:2018rki,Zhu:2019igg,Reiten:2019fta,Zhou:2020ssi,Braga:2020hhs,Arefeva:2020byn,Arefeva:2020vae,Gursoy:2020kjd,Arefeva:2021mag,Arefeva:2021jpa,Zhao:2021ogc,Shahkarami:2021gzl,
Deng:2021kyd,Gursoy:2018ydr,Ali-Akbari:2015bha,Braga:2018zlu,He:2020fdi,Avila:2019pua,Avila:2020ved}. For a related holographic discussion on the free energy and entropy of the heavy quarkonium, see \cite{Iatrakis:2015sua,Dudal:2017max,Ewerz:2016zsx,BitaghsirFadafan:2015zjc,Zhou:2021sdy,Bellantuono:2017msk,Zhang:2020zeo,Asadi:2021nbd}.

The paper is organized as follows. In Section 2, we will first discuss the salient features of the bottom-up magnetised EMD model of \cite{Bohra:2019ebj,Bohra:2020qom}. We will then discuss the free energy, entropy of a quark-antiquark pair, and entropic force in section 3, and compare them with lattice QCD results. Here, we discuss the results for both parallel and perpendicular orientations of the quark-antiquark pair with respect to the magnetic field. The discussion on the dissociation time is presented in Section 4. Finally, we summarize our main results with some discussions and an outlook to future research in Section 5.

\section{Magnetised Einstein-Maxwell-dilaton gravity set up}
Let us first briefly review the magnetised AdS solution of EMD gravity considered in \cite{Bohra:2019ebj,Bohra:2020qom}.  The gravity action is given by,
\begin{eqnarray}
S_{EM} =  -\frac{1}{16 \pi G_5} \int \mathrm{d^5}x \sqrt{-g}  \ \left[R - \frac{f(\phi)}{4}F_{MN}F^{MN} -\frac{1}{2}D_{M}\phi D^{M}\phi -V(\phi)\right]\,,
\label{actionEMD}
\end{eqnarray}
where $R$ is the Ricci scalar, $G_5$ is the Newton's constant in five dimensions, $F_{MN}$ is the field strength tensor of the $U(1)$ gauge field $A_M$, and $L$ is the AdS radius. Moreover, $f(\phi)$ is the gauge kinetic function that acts as coupling between the $U(1)$ gauge field and dilaton field $\phi$, and $V(\phi)$ is the potential of the dilaton field. For more details about this action see \cite{Bohra:2019ebj,Bohra:2020qom}. By varying the action, one can obtain the corresponding Einstein, Maxwell, and dilaton equations of motion. Interestingly, using the following Ans\"{a}tze for the metric field $g_{MN}$, dilaton field $\phi$ and field tensors $F_{MN}$,
\begin{eqnarray}
& & ds^2=\frac{L^2 e^{2A(z)}}{z^2}\biggl[-g(z)dt^2 + \frac{dz^2}{g(z)} + dx_{1}^2+ e^{B^2 z^2} \biggl( dx_{2}^2 + dx_{3}^2 \biggr) \biggr]\,, \nonumber \\
& & \phi=\phi(z), \ \  F_{MN}=B dx_{2}\wedge dx_{3}\,,
\label{ansatze}
\end{eqnarray}
and using appropriate boundary conditions in combination with the potential reconstruction method \cite{Mahapatra:2020wym,Dudal:2018ztm,Priyadarshinee:2021rch,Mahapatra:2018gig,Mahapatra:2019uql,He:2013qq,Arefeva:2018hyo,Arefeva:2018cli,Cai:2012xh}, the EMD equations of motion can be exactly solved in a closed-form in terms of a single form function $A(z)$ \cite{Bohra:2019ebj},
\begin{eqnarray}
g(z) &=& 1 - \frac{ \int_0^z \, d\xi \ \xi^3 e^{-B^2 \xi^2 -3A(\xi) } }{\int_0^{z_h} \, d\xi \ \xi^3 e^{-B^2 \xi^2 -3A(\xi) }} \,, \nonumber \\
\phi(z) &=& \int \, dz \sqrt{-\frac{2}{z} \left(3 z A''(z)-3 z A'(z)^2+6 A'(z)+2 B^4 z^3+2 B^2 z\right)} + K_5 \,, \nonumber \\
f(z) &=& g(z)e^{2 A(z)+2 B^2 z^2} \left(-\frac{6 A'(z)}{z}-4 B^2+\frac{4}{z^2}\right)-\frac{2 e^{2 A(z)+2 B^2 z^2} g'(z)}{z} \,, \nonumber \\
V(z) &=& g'(z) \left(-3 z^2 A'(z)-B^2 z^3+3 z\right) e^{-2 A(z)} - g(z)\left(12 + 9 B^2 z^3 A'(z) \right) e^{-2 A(z)} \nonumber \\
& & +g(z) \left(-9 z^2 A'(z)^2-3 z^2 A''(z)+18 z
   A'(z)-2 B^4 z^4+8 B^2 z^2\right)e^{-2 A(z)} \,,
\label{EMDblackholespl}
\end{eqnarray}
where $z$ is the usual holographic radial coordinate that runs from $z=0$ (asymptotic boundary) to $z=z_h$. This gravity solution corresponds to a black hole with a horizon at $z=z_h$. The background magnetic field is chosen in the $x_1$-direction, which breaks the $SO(3)$ invariance of the boundary spatial coordinates $(x_1, x_2, x_3)$. The constant $K_5$ appearing in Eq.~(\ref{EMDblackholespl}) is fixed by demanding that $\phi |_{z=0}\rightarrow 0$ (asymptotically AdS).

Here we have modeled the magnetic field as a constant external field to gain first insights into the relevant physics. This simplification allows analytical control over most of the computations and is also quite common in magnetised holographic QCD model building. This being said, it has been discussed in \cite{Tuchin:2013ie} that after a fast initial decrease, the magnetic field is more or less frozen for the rest of the lifetime of the plasma, granting credit to the usual assumption of a constant magnetic field.

This black hole solution has the temperature and entropy,
\begin{eqnarray}
& & T = \frac{z_{h}^{3} e^{-3A(z_h)-B^2 z_{h}^{3}}}{4 \pi \int_0^{z_h} \, d\xi \ \xi^3 e^{-B^2 \xi^2 -3A(\xi) } } \,,   \nonumber \\
& & S_{BH} = \frac{V_3 e^{3 A(z_h)+B^2 z_{h}^{3}}}{4 G_5 z_{h}^3 } \,,
\label{BHtemp}
\end{eqnarray}
where $V_3$ is the volume of the three-dimensional spatial plane. For completeness, the magnetic field $B$ in $5D$ actually carries
a mass dimension of one. The physical QCD (boundary) magnetic field, with mass dimension two, will be related to $B$ via a suitable rescaling with $1/L$, but the proportionality constant is undetermined in the current setting. It would require computing an observable in the model and its QCD counterpart and match on top of each other.

There also exists another solution to the EMD equations of motion, corresponding to thermal-AdS (without horizon). This no-black hole solution can be obtained by taking the limit $z_h\rightarrow \infty$, giving $g(z)=1$. The thermal-AdS solution again asymptotes to AdS at the boundary $z=0$, but depending upon the form of $A(z)$, it can have a non-trivial structure in the bulk. Interestingly, there can be a Hawking/Page type phase transition between these two solutions. Interestingly, now the transition temperature becomes a magnetic field dependent quantity. In \cite{Bohra:2019ebj}, the thermal-AdS/black hole phases were shown to be favoured at low/high temperatures respectively, with the transition temperature also found to be decreasing with $B$ (see below for more details). Moreover, these thermal-AdS/black hole phases were shown to be dual to confined/deconfined phases in the boundary theory, thence providing a holographic model for inverse magnetic catalysis in the deconfinement sector.

It is important to stress that Eq.~(\ref{EMDblackholespl}) forms a self-consistent solution of the magnetised EMD gravity for any choice of the form factor $A(z)$. One therefore can construct a large class of analytic solutions of the gravity system (\ref{actionEMD}) by choosing different $A(z)$. Nonetheless, in the context of AdS/QCD, it is reasonable to fix its form by taking inputs from the dual boundary QCD theory. In \cite{Bohra:2019ebj,Bohra:2020qom}, two such forms (i) $A(z)=A_1(z)=-az^2$, and (ii) $A(z)=A_2(z)=-az^2-d B^2 z^5$ were suggested. These two forms were chosen not just for their simplicity, thereby providing better control over the integrals that appear in Eq.~(\ref{EMDblackholespl}), but also for leading to many desirable QCD properties in the boundary theory. A few salient features of these form factors are in order:
\begin{itemize}
\item These forms make sure that the constructed solution asymptotes to AdS at the boundary $z\rightarrow0$.

\item The mass of $\phi$ satisfies the Breitenlohner-Freedman bound for stability in AdS space \cite{Breitenlohner:1982bm}.

\item The dilaton potential is bounded from above by its UV boundary value, thereby satisfying the Gubser stability criterion for a well-defined boundary theory \cite{Gubser:2000nd}.

\item The vector meson mass spectrum exhibits the linear Regge behaviour.

\item The null energy condition of the matter field is always satisfied.

\item Importantly, both forms exhibit a Hawking/Page type phase transition between ther\-mal-AdS and black hole solutions. In particular, thermal-AdS is favoured at low temperatures whereas the black hole phase is favoured at high temperatures. Correspondingly, there is a phase transition between these two. However, since $B$ explicitly appears in various geometric expressions, now the transition temperature is a magnetic field dependent quantity. The behaviour of the transition temperature as a function of the magnetic field is shown in Figures~\ref{TdeconfvsBforA1} and \ref{TdeconfvsBforA2}.

\begin{figure}[t!]
\begin{minipage}[b]{0.5\linewidth}
\centering
\includegraphics[width=2.8in,height=2.3in]{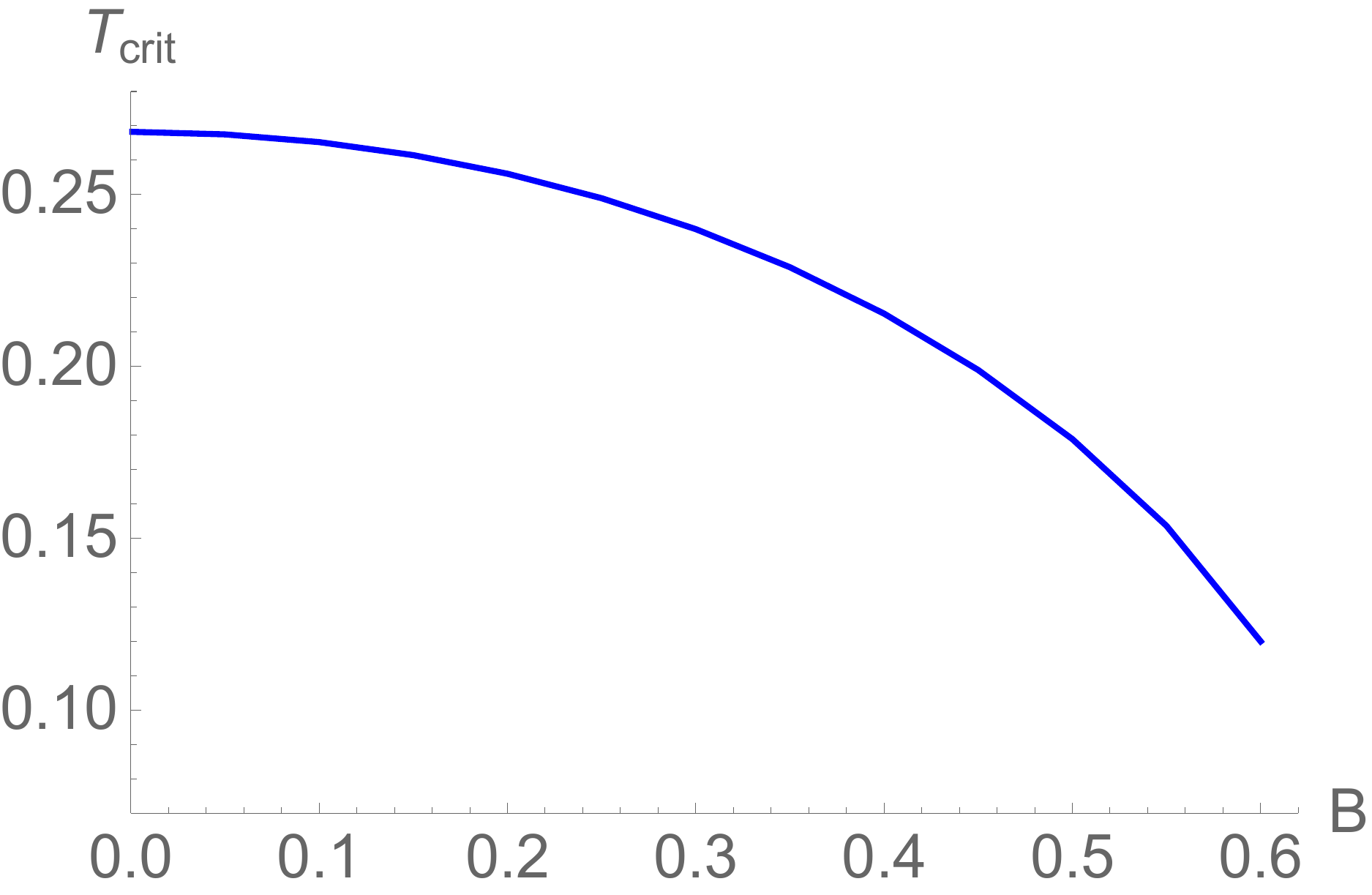}
\caption{ \small Deconfinement transition temperature in terms of magnetic field for the cases $A_1(z)$. Here $a=0.15~\text{GeV}^2$ is used.}
\label{TdeconfvsBforA1}
\end{minipage}
\hspace{0.4cm}
\begin{minipage}[b]{0.5\linewidth}
\centering
\includegraphics[width=2.8in,height=2.5in]{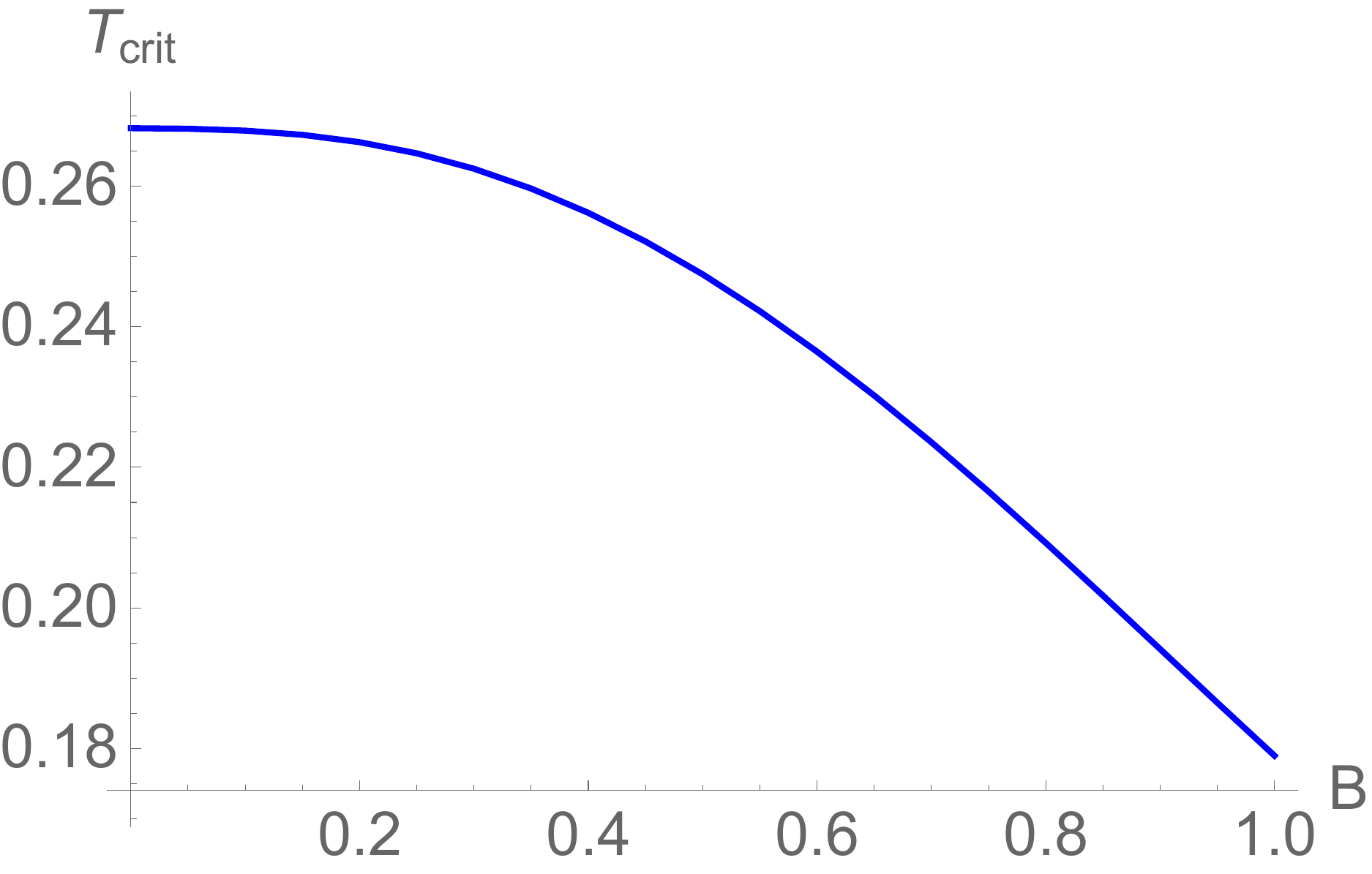}
\caption{\small Deconfinement transition temperature in terms of magnetic field for the cases $A_2(z)$. Here $a=0.15~\text{GeV}^2$ and $d=0.013~\text{GeV}^3$ are used. }
\label{TdeconfvsBforA2}
\end{minipage}
\end{figure}

\item The thermal-AdS and black hole phases are dual to confined and deconfined phases, respectively, in the dual boundary theory. Since the transition temperature decreases with $B$ for both $A(z)$, they both provided a holographic model for inverse magnetic catalysis in the deconfinement sector.

\item The magnitude of the parameter $a$ in both form factors is fixed by demanding the deconfinement transition temperature to be around $270~\text{MeV}$ in the pure glue sector at zero chemical potential and magnetic field. This fixes the value of $a$ to be $a=0.15~\text{GeV}^2$. This also fixes the largest attainable magnitude of the magnetic field, by requiring the real-valuedness of the dilaton field, to be around $B\backsimeq 0.61~\text{GeV}$ for the $A_1(z)$ form factor.

\item The second form factor $A_2(z)=-az^2-d B^2 z^5$ was chosen to make sure that the string breaking does not happen even at a relatively large magnetic field in the confined phase (which is not the case with the first form factor, where the string breaking can happen for $B>0.3~\text{GeV}$ in the confined phase). In particular, with $A_2(z)$, the disconnected quark-antiquark free energy is always larger than the connected quark-antiquark free energy, suggesting no string breaking in the confined phase, as is appropriate for a quenched setup. Similarly, the value of $d$ can be chosen by simultaneously demanding inverse magnetic catalysis, the real-valuedness of the dilaton field, and the largest attainable value for the magnetic field. This fixes $d=0.013~\text{GeV}^3$, see \cite{Bohra:2020qom}.

\item Similarly, in the chiral sector, the magnitude of the chiral condensate increases with $B$ in the confined phase whereas in the deconfinement phase it exhibits a nonmonotonic thermal behaviour for all magnetic field values. The chiral critical temperature again goes down with the magnetic field, indicating inverse magnetic catalysis behaviour. These chiral results also agree qualitatively well with lattice QCD findings, where a similar behaviour has been seen in the chiral sector.
\end{itemize}
We see that with $A_1(z)$ and $A_2(z)$ form factors, the dual boundary theory exhibits many desirable QCD features. Therefore, it is reasonable to continue using these form factors to make predictions (without introducing new free parameters) for the thermal and anisotropic profiles of the quark-antiquark free energy and entropy. Since the $A_2(z)$ form factor minimizes some of the drawbacks of the $A_1(z)$ form factor, while essentially sharing the same QCD properties, it is therefore reasonable to consider only the $A_2(z)$ form factor. Accordingly, in this work we mainly concentrate on the gravity solution based on the $A_2(z)$ form factor. Needless to say, analogous computation can be easily performed for the $A_1(z)$ form factor as well.

\section{Free energy, entropy and entropic force}
The free energy and entropy of the $q\bar{q}$ pair can be computed holographically from the world sheet on-shell action. To do this, one places the $q\bar{q}$ pair with interquark distance $\ell$ on the asymptotic AdS boundary with a string that stretches between them and lets it evolve with time. The free energy $\mathcal{F}$ of the pair is then given by the two-dimensional area swept out by the fundamental string, with the boundary condition that the two-dimensional world-sheet is bounded by the rectangular Wilson loop of sides $\ell\times T$ at the AdS boundary \cite{Maldacena:1998im,Brandhuber:1998bs,Rey:1998ik}. We have
\begin{eqnarray}
\mathcal{F}(T,\ell)=T \ S_{NG}^{on-shell} \,,
\label{Sonshell}
\end{eqnarray}
where $T$ is the temperature and $S_{NG}^{on-shell}$ is the on-shell Nambu-Goto action,
\begin{eqnarray}
S_{NG}=\frac{1}{2 \pi \ell_{s}^2}\int d\tau d\sigma \sqrt{-\det g_s}, \ \ \ (g_s)_{\alpha\beta}=(g_s)_{MN}\partial_\alpha X^{M} \partial_\beta  X^{N} \,.
\label{NGaction}
\end{eqnarray}
Here, $T_s=1/2 \pi \ell_{s}^2$ is the open string tension, $(\tau,\sigma)$ are the world sheet coordinates, and $X^{M}(\tau,\sigma)$ denotes the open string embedding. $g_s$ denotes the background metric in string frame~\footnote{Henceforth,  a subscript ``$s$" is used to denote quantities in the string frame.}. It is related to the Einstein frame metric of the previous section in the following way,
\begin{eqnarray}
& & (g_s)_{MN}=e^{\sqrt{\frac{2}{3}}\phi} g_{MN} \,,  \nonumber \\
& & ds_{s}^2= \frac{L^2 e^{2 A_{s}(z)}}{z^2}\biggl[-g(z)dt^2 + \frac{dz^2}{g(z)} + dx_{1}^2+ e^{B^2 z^2} \left(dx_{2}^2+dx_{3}^2\right) \biggr] \,,
\label{stringmetric}
\end{eqnarray}
where $A_{s}(z)=A(z)+\sqrt{\frac{1}{6}} \phi(z)$.

To proceed further, we need to choose the direction of $q\bar{q}$ pair and appropriately parameterize the string. Importantly, with a background magnetic field, two interesting scenarios are possible: i) $q\bar{q}$ pair oriented parallel to the magnetic field and ii) $q\bar{q}$ pair oriented perpendicular to the magnetic field.

\subsection{Parallel case}
In this case, we take $\sigma=x_1$ to parameterize the string world-sheet and work in the static gauge $\tau=t$. The string is embedded in the five-dimensional bulk spacetime, i.e., $z=z(x_1)$. This gives us,
\begin{eqnarray}
\mathcal{F}^{\parallel}(T,\ell) = \frac{L^2}{2 \pi \ell_{s}^2} \int\, dz \frac{e^{2A_s(z)}}{z^2} \sqrt{1+ \frac{g(z)}{z'^2}}\,
\end{eqnarray}
where prime $'$ denotes a derivative with respect to $x_1$. The endpoints of the string are located at $z(-\ell/2)=z(\ell/2)=0$. It turns out that there are two string world sheet configurations that minimize the free energy: a connected and a disconnected one. The connected string is a $\cup$-shape configuration that stretches from the boundary into the bulk whereas the disconnected string configuration consists of two lines that are separated by a distance $\ell$ and are stretched from the boundary to the horizon. In the holographic language, the connected string corresponds to a bound $q\bar{q}$ state whereas the disconnected string corresponds to a dissociated quarkonium state. For the connected case, the free energy expression reduces to
\begin{equation}
\mathcal{F}_{con}^{\parallel}=\frac{L^{2}}{ \pi l_{s}^{2}}\int_{0}^{z_{*}^{\parallel}}\,dz \frac{z_{*}^{\parallel 2}}{z^{2}} \frac{\sqrt{g(z)}e^{2A_{s}(z)-2A_{s}(z_{*}^{\parallel})}}{\sqrt{g(z)z_{*}^{\parallel 4}e^{-4A_{s}(z_{*}^{\parallel})}-g(z_{*}^{\parallel})z^{4}e^{-4 A_{s}(z)}}} \,,
\label{freeEnergyconnparallel}
\end{equation}
where $z_{*}^{\parallel}$ is the turning point of the connected string. In particular, the world sheet descends into the bulk geometry from the boundary, reaching a turning point $z=z^{\parallel}_*$ at $x_1=0$, before symmetrically ascending again towards the asymptotic boundary. The turning point $z=z^{\parallel}_*$ is related to the separation length $\ell^{\parallel}$ of the $q\bar{q}$ pair as
\begin{equation}
\ell^{\parallel}=2\int_{0}^{z_{*}^{\parallel}}\,dz \sqrt{\frac{g(z_{*}^{\parallel})}{g(z)}}\frac{z^{2}e^{-2A_{s}(z)}}{\sqrt{g(z)z_{*}^{\parallel 4}e^{-4A_{s}(z_{*}^{\parallel})}-g(z_{*}^{\parallel})z^{4}e^{-4 A_{s}(z)}}} \,.
\label{lengthparallel}
\end{equation}
Similarly, the free energy expression of the disconnected string reduces to
\begin{equation}
\mathcal{F}_{discon}^{\parallel}=\frac{L^{2}}{\pi l_{s}^{2}} \int_{0}^{z_h}\, dz\frac{e^{2A_{s}(z)}}{z^{2}}\,.
\end{equation}
Notice that $\mathcal{F}_{discon}^{\parallel}$ is independent of $z_{*}^{\parallel}$ and therefore of the interquark distance $\ell^{\parallel}$. The $\mathcal{F}_{discon}^{\parallel}$ can be thought of as the free energy of the non-interacting heavy quarks. Expectedly, both $\mathcal{F}_{con}^{\parallel}$ and $\mathcal{F}_{discon}^{\parallel}$ contain UV divergences in the limit $z\rightarrow 0$. However, the nature of the poles is the same in both $\mathcal{F}_{con}^{\parallel}$ and $\mathcal{F}_{discon}^{\parallel}$. Therefore, these divergences trivially cancel out in the free energy difference $\Delta\mathcal{F}^{\parallel}=\mathcal{F}_{con}^{\parallel}-\mathcal{F}_{discon}^{\parallel}$, which is our main quantity of interest.  Also, where-ever required, we will use the temperature-independent minimum renormalization scheme of \cite{Ewerz:2016zsx}, which corresponds to minimally subtracting the poles when cutting the integral at $z=\varepsilon_{UV}\ll1$. \footnote{In \cite{Ewerz:2016zsx}, $\Delta\mathcal{F}^{\parallel}$ has been suggested as the binding energy of the $q\bar{q}$ pair for small interquark separations.}

\begin{figure}[h!]
\begin{minipage}[b]{0.5\linewidth}
\centering
\includegraphics[width=2.8in,height=2.3in]{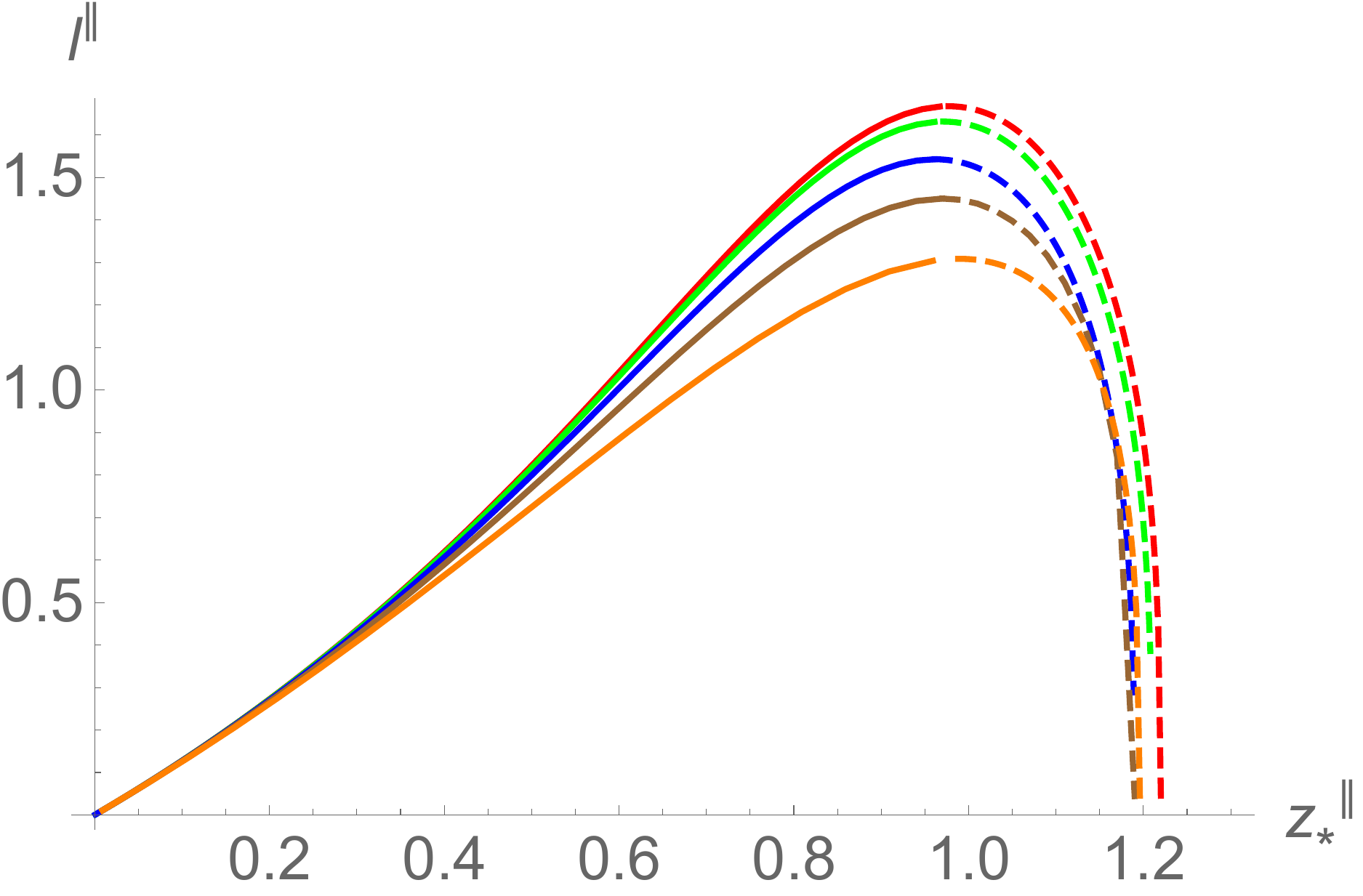}
\caption{ \small $\ell^{\parallel}$ as a function of $z_{*}^{\parallel}$ for various values of magnetic field. Here $T=1.2~T_{crit}$ is used. Red, green, blue, brown, and orange
curves correspond to $B=0$, $0.2$, $0.4$, $0.6$, and $0.8$ respectively. In units of GeV.}
\label{zsvslvsBT1Pt2TcparallelforA1}
\end{minipage}
\hspace{0.4cm}
\begin{minipage}[b]{0.5\linewidth}
\centering
\includegraphics[width=2.8in,height=2.5in]{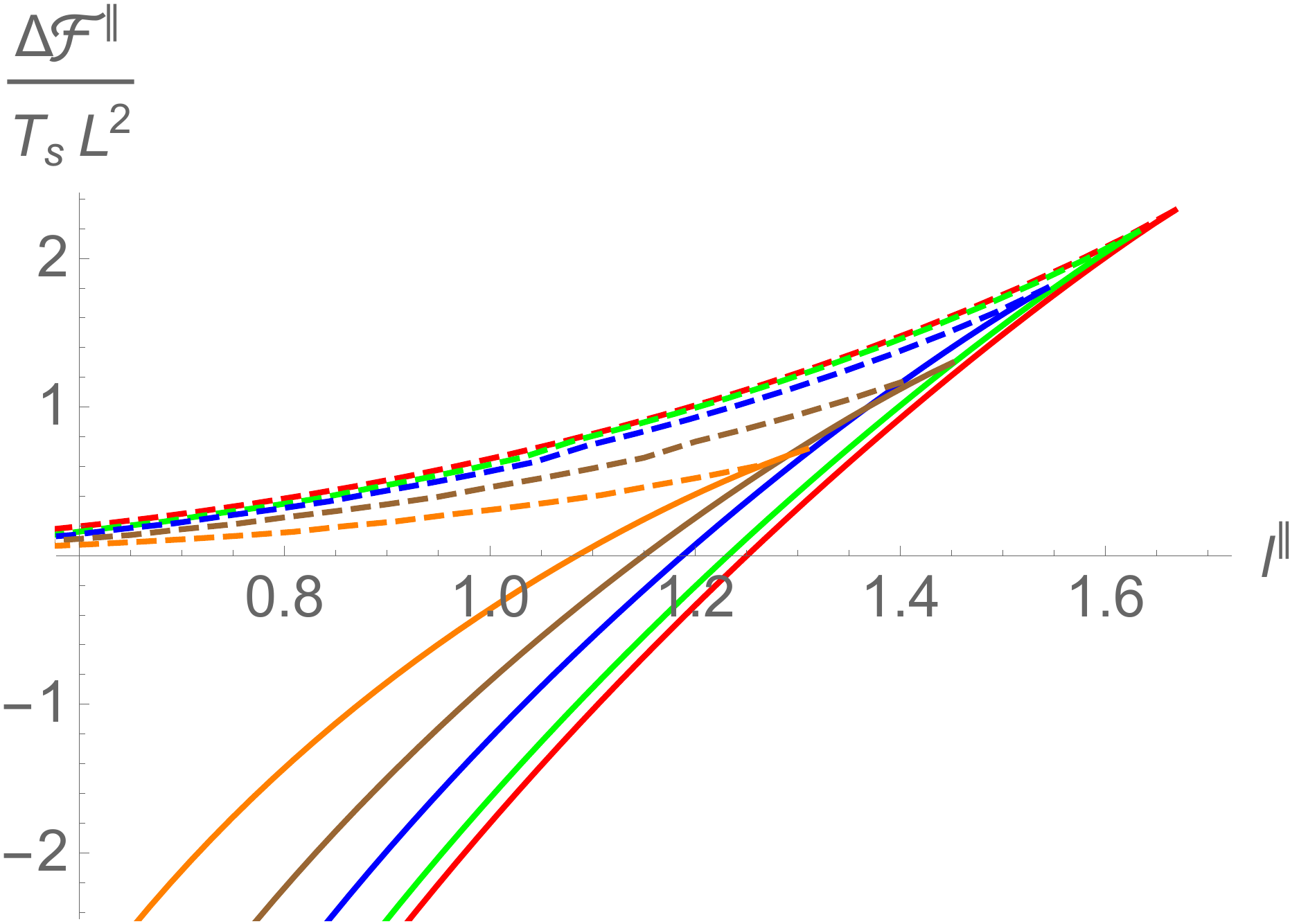}
\caption{\small $\Delta\mathcal{F}^{\parallel}$ as a function of $\ell^{\parallel}$ for various values of magnetic field. Here $T=1.2~T_{crit}$ is used. Red, green, blue, brown, and orange
curves correspond to $B=0$, $0.2$, $0.4$, $0.6$, and $0.8$ respectively. In units of GeV.}
\label{lvsDeltaFvsBT1Pt2TcparallelforA1}
\end{minipage}
\end{figure}

Let us first discuss the numerical results for different magnetic fields. The behaviour of $\ell^{\parallel}$ with respect to the connected string turning point $z_{*}^{\parallel}$ is illustrated in Figure~\ref{zsvslvsBT1Pt2TcparallelforA1}. Here, we have taken $T=1.2~T_{crit}$ for simplicity but similar results appear for other temperatures as well. We observe that at each magnetic field there appears a maximum separation length $\ell_{max}$ above which the connected string solution does not exist, and only the disconnected solution remains. However, below $\ell_{max}$, there appear two $z_{*}^{\parallel}$ solutions at each $\ell^{\parallel}$ for the connected string. The one solution corresponding to small $z_{*}^{\parallel}$ (solid lines) is closer to the boundary compared to the second  large $z_{*}^{\parallel}$ solution (dashed lines). It turns out that (as discussed below) the former solution is actually the true minimum whereas the latter solution corresponds to a saddle point. As $\ell^{\parallel}$ grows, the two connected solutions approach each other before merging into a single solution at $\ell^{\parallel}_{max}$.  Importantly, none of the connected solutions touch the horizon. This in a way associates a length scale with $\ell^{\parallel}_{max}$. In particular, $\ell^{\parallel}_{max}$ can be thought of as the maximally possible separation length at which the $q\bar{q}$ form a bound state connected by a string \cite{Ewerz:2016zsx}. For larger separations, the quark and antiquark are screened by the thermal medium.

The free energy difference $\Delta\mathcal{F}^{\parallel}$ between connected and disconnected strings is shown in Figure~\ref{lvsDeltaFvsBT1Pt2TcparallelforA1}. Here, solid and dashed lines again correspond to small and large $z_{*}^{\parallel}$ solutions respectively. We observe that the small $z_{*}^{\parallel}$ solution always has a lower free energy than the large $z_{*}^{\parallel}$ solution, validating our earlier statement. Importantly, there occurs a phase transition between connected and disconnected strings as the interquark separation length is varied. In particular, $\Delta\mathcal{F}^{\parallel}$ changes sign from negative to positive as $\ell^{\parallel}$ is increased, implying that the connected string has a lower free energy at small separations whereas the disconnected string has a lower free energy at large separations. Since $\mathcal{F}_{discon}^{\parallel}$ is independent of $\ell^{\parallel}$, the corresponding QCD string tension is zero. It suggests that there is no linear confinement law in the boundary theory dual to the black hole. Moreover, the length at which the connected/disconnected phase transition occurs defines a critical length $\ell^{\parallel}_{crit}$ ($<\ell^\parallel_{max}$), i.e., at $\ell^{\parallel}_{crit}$ the free energy of the $q\bar{q}$ pair exhibits a non-smooth behaviour. We find that $\ell^{\parallel}_{crit}$ decreases with the magnetic field near the deconfinement temperature. This implies that quarkonium dissociation, which is described by the disconnected string configuration in the dual gravity picture, gets promoted by the magnetic field. Put in other words, the size of the heavy quark bound state decreases with a parallel magnetic field near the deconfinement temperature. However, away from this temperature, $\ell^{\parallel}_{crit}$ seems to increase slightly with the magnetic field. Nonetheless, we expect that at extremely large temperatures the differences due to a magnetic field will be negligible, as thermal effects will eventually dominate the physics. Indeed, for $T=10~T_{crit}$, the magnitude of $\ell^{\parallel}_{crit}$ at  $B=1~\text{GeV}$ differs from its $B=0$ counterpart only at the third decimal place.

\begin{figure}[h!]
\centering
\includegraphics[width=2.8in,height=2.3in]{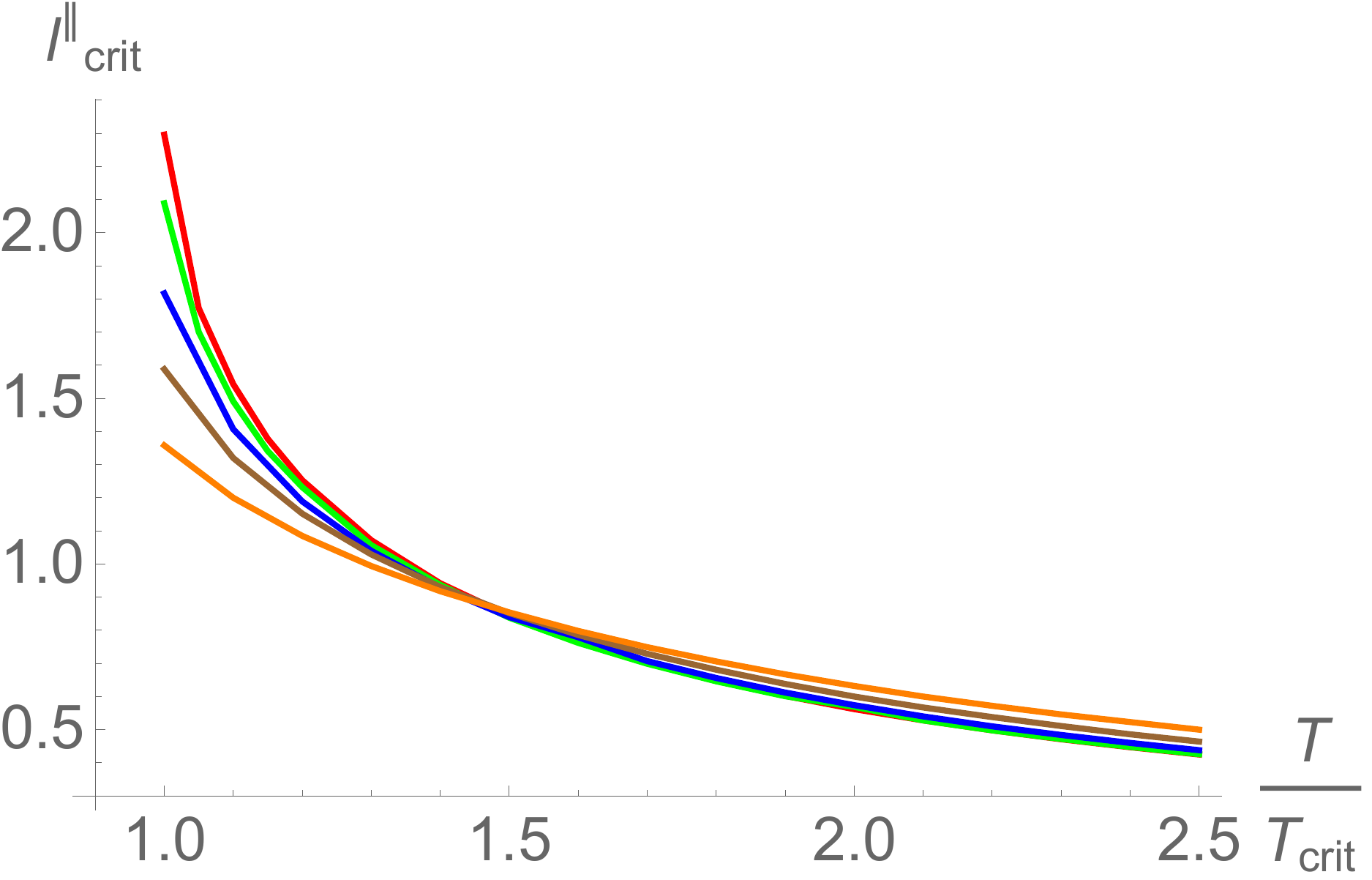}
\caption{ \small $\ell^{\parallel}_{crit}$ as a function of temperature for various values of magnetic field. Red, green, blue, brown, and orange curves correspond to $B=0$, $0.2$, $0.4$, $0.6$, and $0.8$ respectively. In units of GeV.}
\label{TvslcritvsBparallelforA1}
\end{figure}

Similarly, we analyse the thermal behaviour of the $q\bar{q}$ free energy. We find that $\ell^{\parallel}_{crit}$ decreases with temperature as well. This structure is consistent with the physical expectation that at larger and larger temperatures the boundary meson state (connected string) would eventually disintegrate to a free quark and antiquark (disconnected string). This is true for all magnetic field values. The complete phase diagram illustrating the dependence of $\ell^{\parallel}_{crit}$ on temperature and magnetic field is shown in Figure~\ref{TvslcritvsBparallelforA1}.

To further appreciate the order of $\ell^{\parallel}_{crit}$ in Figure~\ref{TvslcritvsBparallelforA1}, let us remind here that the typical binding size of heavy quark bound states, like charmonia, is of order $0.5~fm\backsimeq 2.5~\text{GeV}^{-1}$. The $\ell^{\parallel}_{crit}$ values shown in Figure~\ref{TvslcritvsBparallelforA1} are not too far from this value. For example, we got $\ell^{\parallel}_{crit}\backsimeq2.3~\text{GeV}^{-1}$ at $B=0$ near the deconfinement temperature. This suggests that $\ell^{\parallel}_{crit}$ values obtained here are reasonably well compatible with QCD phenomenological relevant length scales. The prediction that $\ell^{\parallel}_{crit}$ decreases appreciatively with $B$ near the deconfinement temperature is an important prediction of our model and could be tested in independent lattice settings.

\begin{figure}[h!]
\centering
\includegraphics[width=2.8in,height=2.3in]{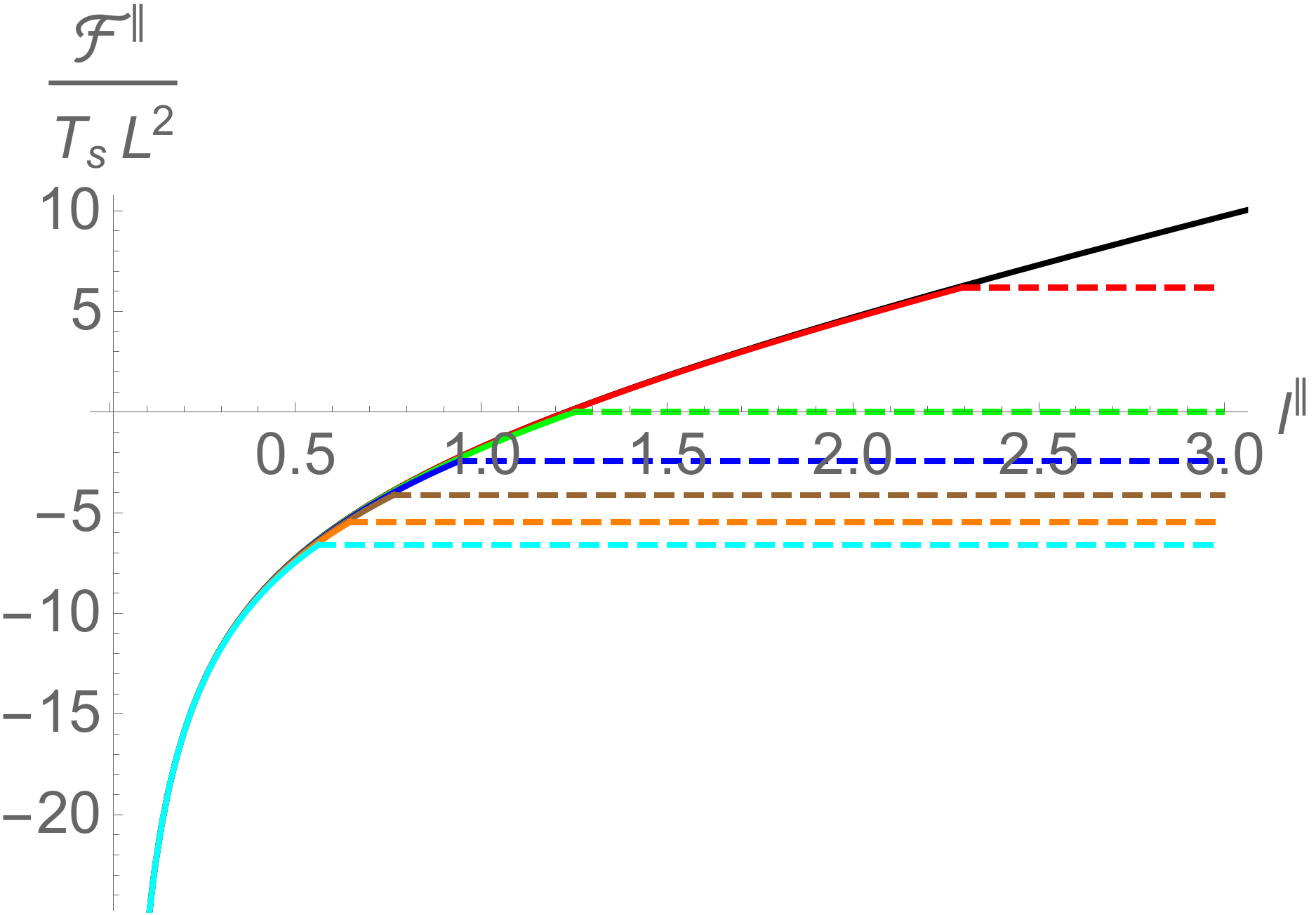}
\caption{ \small $\mathcal{F}^{\parallel}$ as a function of $\ell^{\parallel}$ for various values of temperature. Here $B=0$ is used. Black, red, green, blue, brown, orange, and cyan
curves correspond to $T/T_{crit}=0$, $1.0$, $1.2$, $1.4$, $1.6$, $1.8$, and $2.0$ respectively. In units of GeV.}
\label{lvsFvsTB0parallelforA1}
\end{figure}

It is also important to explicitly analyse the $\ell^{\parallel}$ dependence of the $q\bar{q}$ free energy for varying temperatures. This is shown in Figure~\ref{lvsFvsTB0parallelforA1} for $B=0$. For completion, its $T=0$ profile is also included \footnote{To obtain the $q\bar{q}$ free energy at $T=0$, we use the thermal-AdS background corresponding to $g(z)=1$.}. For small separations $\ell^{\parallel}T\ll1$, the free energy turns out to be independent of temperature. The temperature independence of the $q\bar{q}$ free energy at small separation is physically expected, as the structure of the UV region of small separations is expected to be not influenced by the thermal scale, i.e., the wavelength of the thermal excitation is of order $1/T$, and hence is expected to be not able to probe the interaction between the $q\bar{q}$ at small separations. This behaviour is also supported by lattice data \cite{Kaczmarek:2002mc,Petreczky:2004pz}, where indeed the $q\bar{q}$ free energy was found to be independent of temperature for small separation. On the gravity side, this corresponds to the fact that for small $\ell^{\parallel}$, the string world sheet does not penetrate the bulk far from the boundary, and accordingly, its propagation is limited and inherently fixed by the boundary conditions alone. Therefore, the parameter temperature which manifests itself explicitly only deeper in the bulk does not modify the string dynamics (and therefore the corresponding $q\bar{q}$ free energy) at small separations.

\begin{figure}[h!]
\begin{minipage}[b]{0.5\linewidth}
\centering
\includegraphics[width=2.8in,height=2.3in]{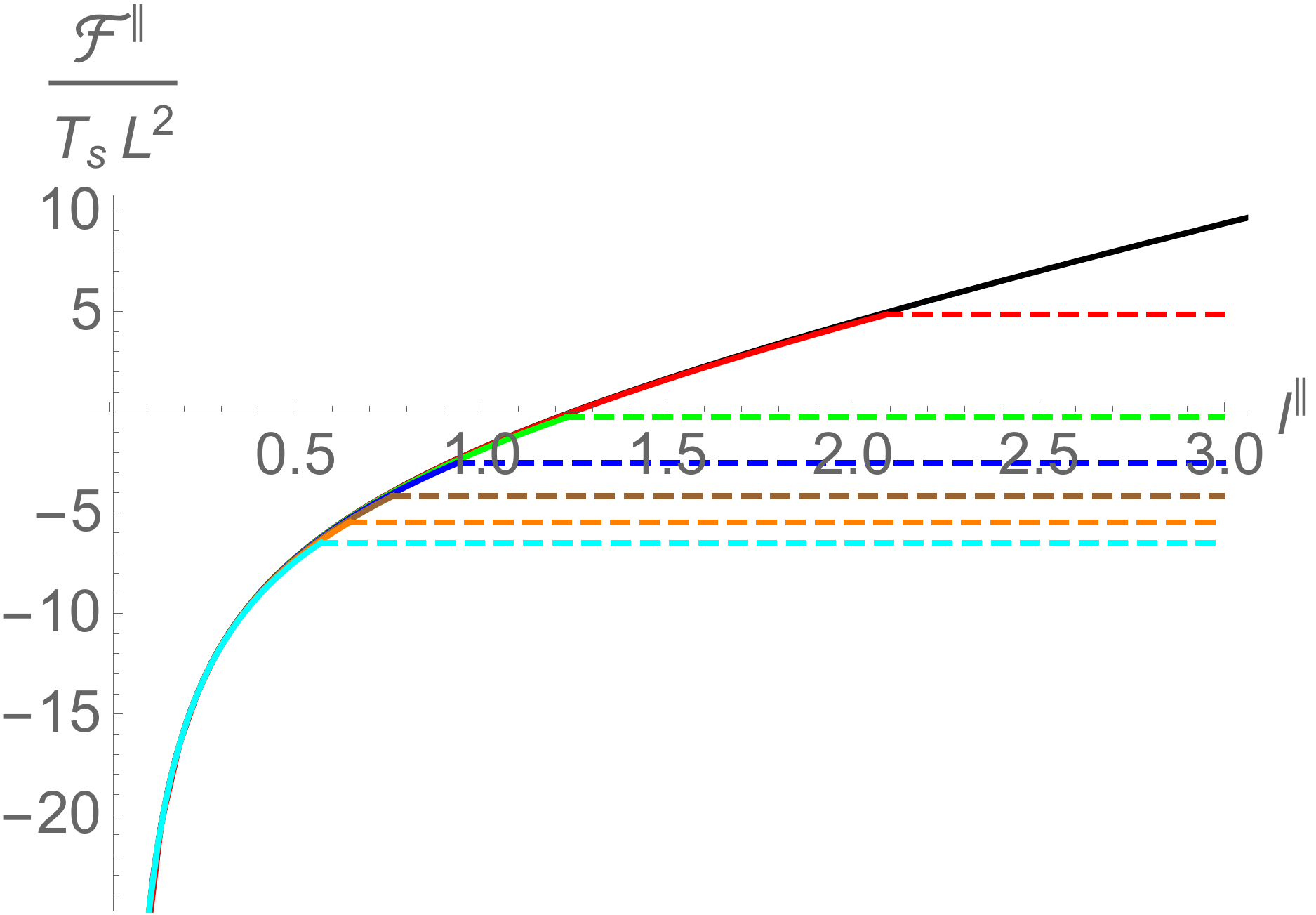}
\caption{ \small $\mathcal{F}^{\parallel}$ as a function of $\ell^{\parallel}$ for various values of temperature. Here $B=0.2$ is used. Black, red, green, blue, brown, orange, and cyan
curves correspond to $T/T_{crit}=0$, $1.0$, $1.2$, $1.4$, $1.6$, $1.8$, and $2.0$ respectively. In units of GeV.}
\label{lvsFvsTBPt2parallelforA1}
\end{minipage}
\hspace{0.4cm}
\begin{minipage}[b]{0.5\linewidth}
\centering
\includegraphics[width=2.8in,height=2.5in]{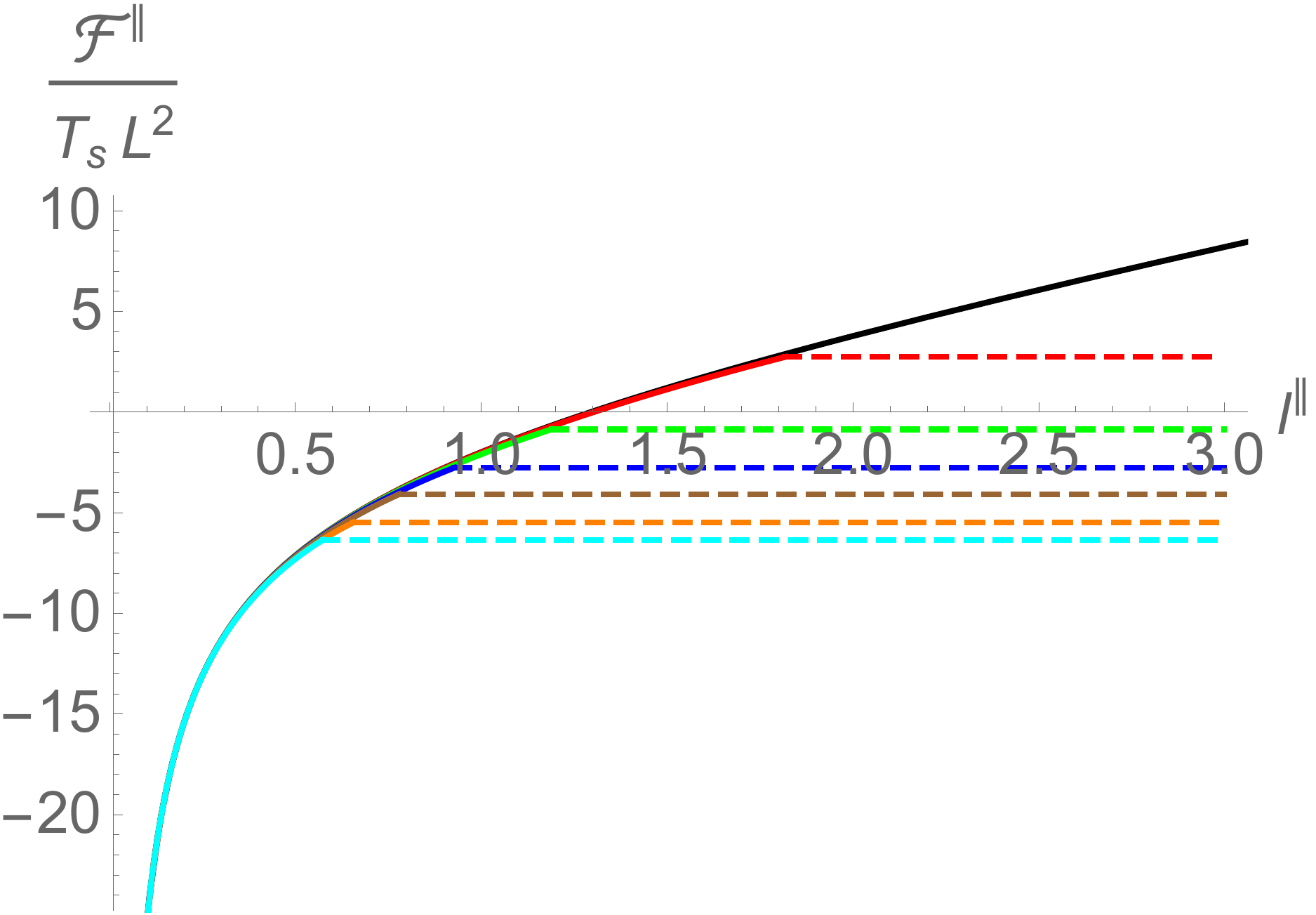}
\caption{\small $\mathcal{F}^{\parallel}$ as a function of $\ell^{\parallel}$ for various values of temperature. Here $B=0.4$ is used. Black, red, green, blue, brown, orange, and cyan
curves correspond to $T/T_{crit}=0$, $1.0$, $1.2$, $1.4$, $1.6$, $1.8$, and $2.0$ respectively. In units of GeV.}
\label{lvsFvsTBPt4parallelforA1}
\end{minipage}
\end{figure}

We further find that the $q\bar{q}$ free energy exhibits a Coulombic behaviour for small $\ell^{\parallel}$ for all values of $T$, i.e., $\mathcal{F}^{\parallel} \propto 1/\ell^{\parallel}$. As $\ell^{\parallel}$ increases, the free energy increases and then saturates to a constant value. The free energy is further found to be decreasing with temperature. In particular, for increasing $T$, $\mathcal{F}^{\parallel}$ at fixed $\ell^{\parallel}$ becomes smaller (as compared to the $T=0$ case). This is expected considering that thermal screening of color charges is expected to be stronger at higher temperatures. These features have also been observed in lattice studies. Moreover, lattice QCD also predicts a flattening of the free energy at large separation. This behaviour is also exhibited by our holographic model. The non-smooth behaviour arises due to the first-order phase transition between connected and disconnected free energies.

The free energy displays a similar behaviour for the finite magnetic field as well. In particular, it not only exhibits a Coulombic structure for small $\ell^{\parallel}$ for all $T$ and $B$ but also increases and then flattens out at large separations. This is shown in Figures~\ref{lvsFvsTBPt2parallelforA1} and \ref{lvsFvsTBPt4parallelforA1}. Moreover, the free energy is again found to be decreasing with temperature for all $B$. Interestingly, the saturation value of the free energy at large separation decreases with the magnetic field for temperatures near the deconfinement temperature whereas it increases with magnetic field at much higher temperatures.

Let us now discuss the effects of the magnetic field on the entropy of the $q\bar{q}$ pair. The entropy can be calculated from the free energy via,
\begin{equation}
S^{\parallel}=-\frac{\partial \mathcal{F}^{\parallel}}{\partial T}\,.
\end{equation}
However, now we have two distinct behaviours of $S$ depending on whether $\ell^{\parallel}<\ell^{\parallel}_{crit}$ or $\ell^{\parallel}>\ell^{\parallel}_{crit}$. In particular, we have
\begin{equation}
S^{\parallel}_{con}(\ell^{\parallel}<\ell^{\parallel}_{crit})=-\frac{\partial \mathcal{F}^{\parallel}_{con}}{\partial T}\,,
\label{Sparallelcon}
\end{equation}
for small interquark separations. Whereas, we have
\begin{equation}
S^{\parallel}_{discon}(\ell^{\parallel}>\ell^{\parallel}_{crit})=-\frac{\partial\mathcal{F}^{\parallel}_{discon}}{\partial T}\,,
\label{Sparalleldiscon}
\end{equation}
for large separations. Interestingly, as is discussed below, the two distinct behaviour of $\ell^{\parallel}$ vs. $\mathcal{F}^{\parallel}$ are important in capturing lattice QCD like results for the entropy from holography.

\begin{figure}[h!]
\centering
\includegraphics[width=2.8in,height=2.3in]{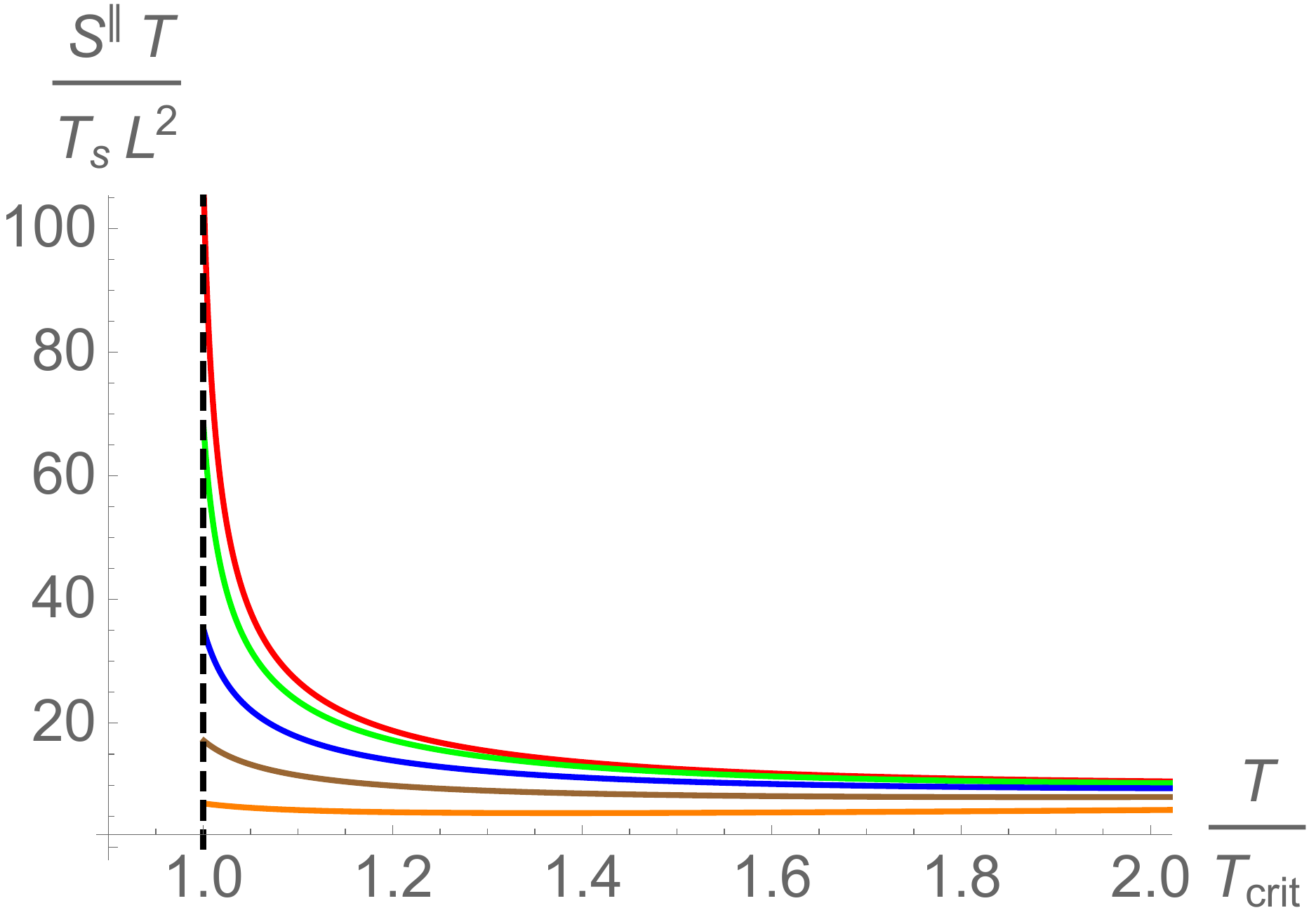}
\caption{ \small $S^{\parallel}$ as a function of temperature for various values of magnetic field. Red, green, blue, brown, and orange curves correspond to $B=0$, $0.2$, $0.4$, $0.6$, and $0.8$ respectively. In units of GeV.}
\label{TvsSvsBlargeLparallelforA1}
\end{figure}

In Figure~\ref{TvsSvsBlargeLparallelforA1}, the $q\bar{q}$ entropy as a function of temperature for various values of the magnetic field at large interquark separation is shown. For this purpose, we have used Eq.~(\ref{Sparalleldiscon}). Our results can be compared to Figure~\ref{latticeQCD1} (cf.~the lattice data of \cite{Kaczmarek:2005zp}). In the confined phase ($T<T_{crit}$), the entropy is identically zero. This is due to the fact that the confining phase is dual to the thermal-AdS background, i.e., without horizon and temperature. This is a generic feature of all large $N_c$ holographic QCD models, and our model is no exception to that. In the deconfinement phase (dual to a black hole), on the other hand, a large amount of entropy is found to be associated with the $q\bar{q}$ pair, especially near the deconfinement temperature. In particular, there is a considerable peak in the $q\bar{q}$ entropy near the deconfinement temperature. These results agree with the lattice results qualitatively.

Interestingly, our analysis further suggests that, compared to the zero magnetic field case, the magnitude of the $q\bar{q}$ entropy at the peak substantially decreases with the finite magnetic field. In particular, $S^{\parallel}$ at $B=0$ is of order magnitude higher than at $B=1~\text{GeV}$ near the deconfinement temperature. This is an important prediction of our model that can be tested in lattice calculations  (as one will not have to worry about various numerical issues, like the sign problem, with a finite magnetic field). Moreover, the higher temperature asymptotic behaviour of the $q\bar{q}$ entropy in our model also mimics lattice QCD. Specifically, for $T\gtrsim 2T_{crit}$, $T S^{\parallel}$ tends to rise with temperature, as is also observed in lattice QCD. Our analysis further predicts similar large temperature asymptotics of $T S^{\parallel}$ with the magnetic field.

\begin{figure}[h!]
\begin{minipage}[b]{0.5\linewidth}
\centering
\includegraphics[width=2.8in,height=2.3in]{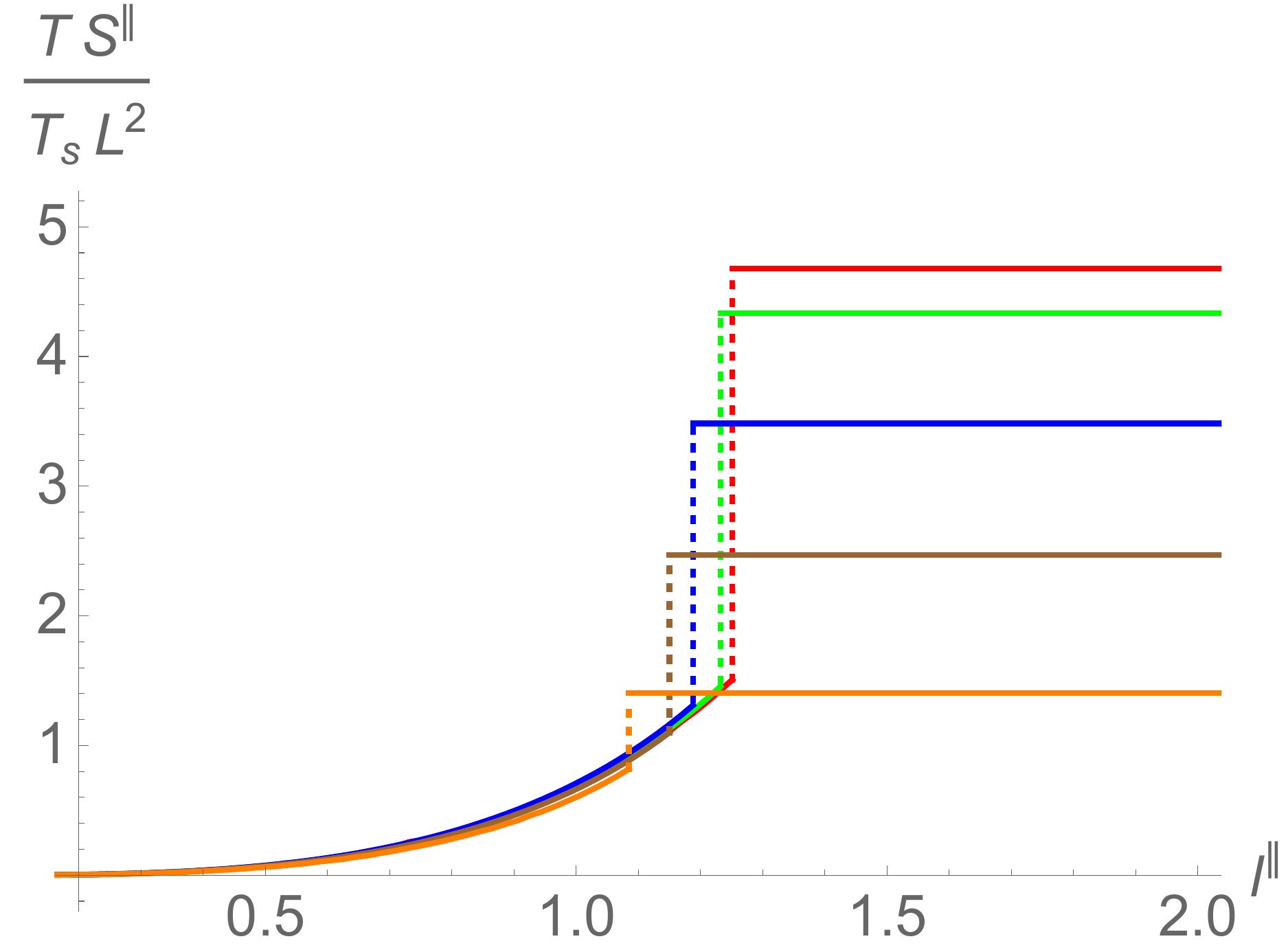}
\caption{ \small $S^{\parallel}$ as a function of $\ell^{\parallel}$ for various values of magnetic field. Here $T=1.2~T_{crit}$ is used. Red, green, blue, brown, and orange curves correspond to $B=0$, $0.2$, $0.4$, $0.6$, and $0.8$ respectively. In units of GeV.}
\label{lvsSvsBT1Pt2TcparallelforA1}
\end{minipage}
\hspace{0.4cm}
\begin{minipage}[b]{0.5\linewidth}
\centering
\includegraphics[width=2.8in,height=2.5in]{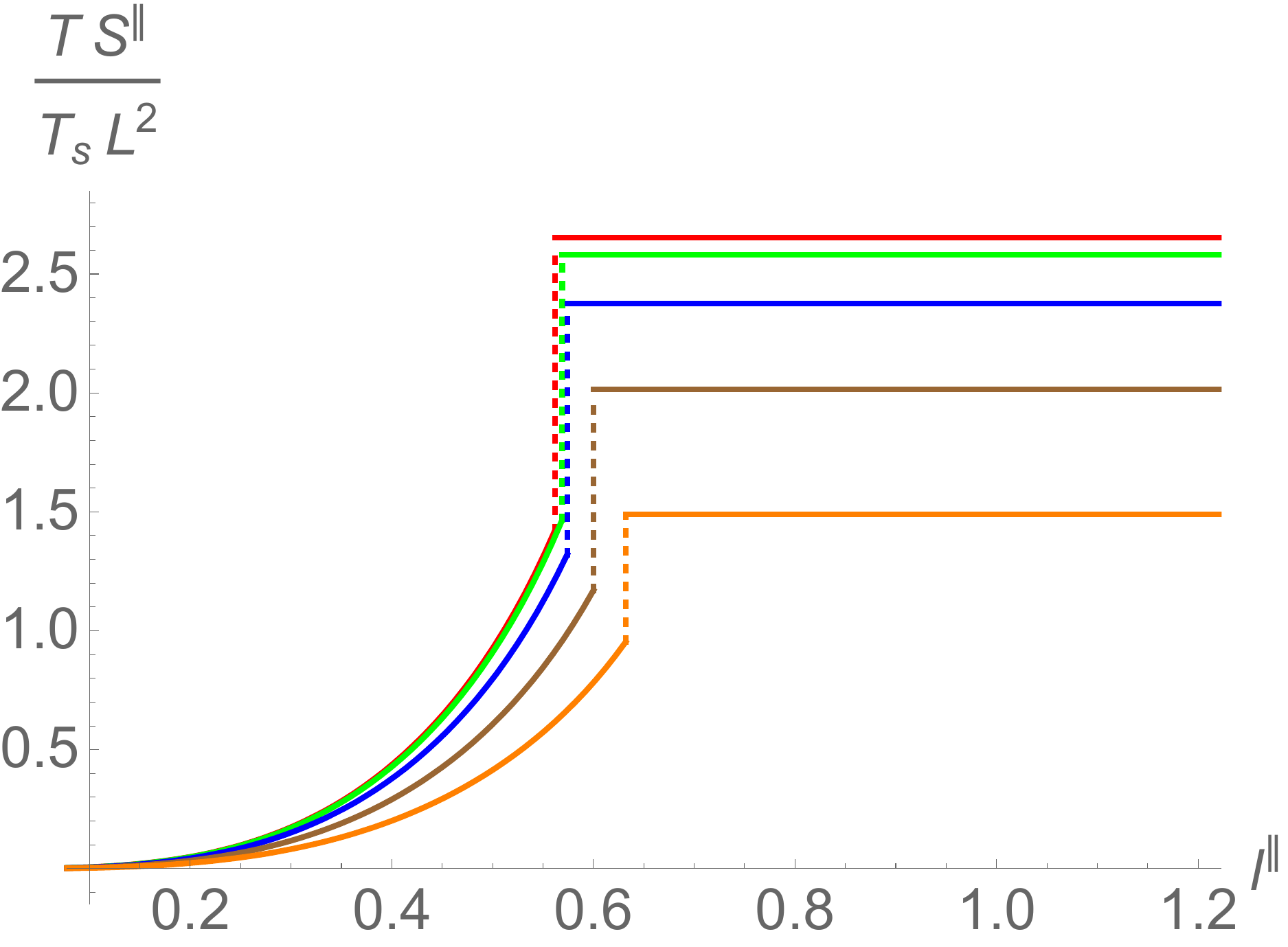}
\caption{\small $S^{\parallel}$ as a function of $\ell^{\parallel}$ for various values of magnetic field. Here $T=2.0~T_{crit}$ is used. Red, green, blue, brown, and orange curves correspond to $B=0$, $0.2$, $0.4$, $0.6$, and $0.8$ respectively. In units of GeV.}
\label{lvsSvsBT2TcparallelforA1}
\end{minipage}
\end{figure}

Another interesting lattice QCD feature that our holographic QCD model reproduces reasonably well is the prediction of an increase in the entropy of $q\bar{q}$ pair with interquark distance. This is illustrated in Figures~\ref{lvsSvsBT1Pt2TcparallelforA1} and \ref{lvsSvsBT2TcparallelforA1}, where $\ell^{\parallel}$ vs. $T S^{\parallel}$ behaviour for various values of the magnetic field at two different temperatures is shown \footnote{To make the figures more readable, the magnitude of entropy $S^{\parallel}$ has been reduced by a factor of 4 in  $\ell > \ell^{\parallel}_{crit}$ region of Figures~\ref{lvsSvsBT1Pt2TcparallelforA1} and \ref{lvsSvsBT2TcparallelforA1}.}. We observe that for all cases, $T S^{\parallel}$ first increases with $\ell^{\parallel}$ and then saturates to a constant value. The $\ell^{\parallel}$ independence of the entropy at large separations arises due to the fact that the relevant disconnected free energy is independent of $\ell^{\parallel}$. We see that these features match qualitatively well again with lattice results (shown in Figure~\ref{latticeQCD2}). However, unlike lattice QCD, the entropy unfortunately approaches saturation value in a discontinuous manner. In particular, there appears a discontinuity in the entropy spectrum at $\ell^{\parallel}_{crit}$. This discontinuity again arises precisely from the first-order transition between connected and disconnected free energies at $\ell^{\parallel}_{crit}$. Another interesting observation is that with the magnetic field the saturation value of the entropy at large interquark distances decreases. We analyze this behaviour untill $T\simeq 3~T_{crit}$, and find it to be true for all temperatures.

\subsection{Entropic force}
\begin{figure}[h!]
	\centering
	\includegraphics[width=2.8in,height=2.3in]{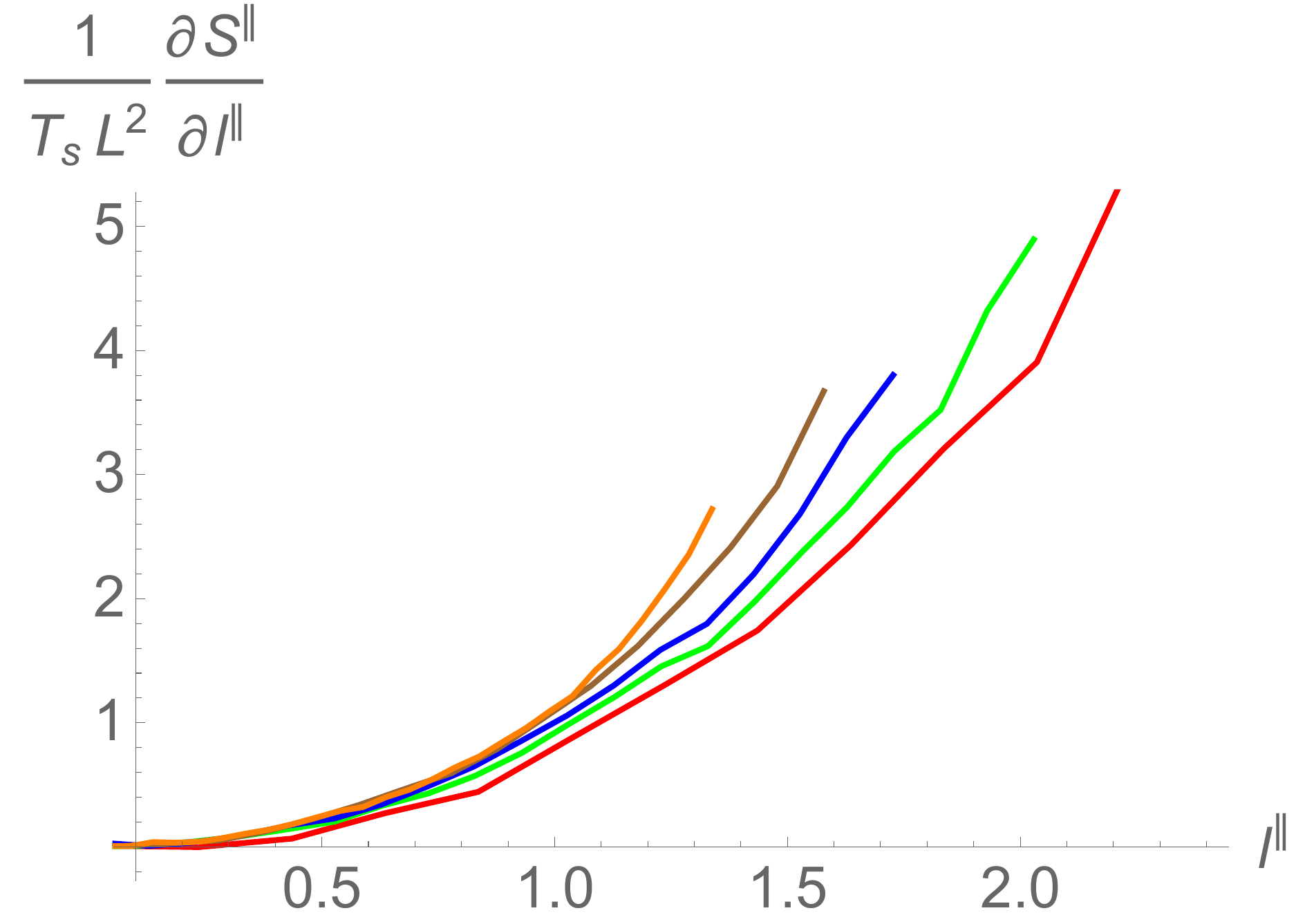}
	\caption{ \small Entropic force  as a function of $\ell^{\parallel}$ for various values of magnetic field in the parallel case. Here $T=1.0~T_{crit}$ is used. Red, green, blue, brown, and orange curves correspond to $B=0$, $0.2$, $0.4$, $0.6$, and $0.8$ respectively. In units of GeV.}
	\label{lvsEntropicforcevsBT1TcparallelforA1}
\end{figure}

From the entropy, we can further compute the entropy force [Eq.~(\ref{entropicforce})]. We find that near the deconfinement temperature, the slope of $\ell^{\parallel}$ vs.~$S^{\parallel}$ is positive and increases with the magnetic field, thereby suggesting an enhancement of the entropic force in the presence of a magnetic field. This is illustrated in Figure~\ref{lvsEntropicforcevsBT1TcparallelforA1}. In the language of \cite{Kharzeev:2014pha}, the strength of the entropic force would promote the dissociation of the heavy quark pair. This finding of the destructive effects of the magnetic field on the heavy quark dissociation once again point towards an inverse response of the QCD system to a magnetic field.

\subsection{Perpendicular case}
We now discuss the free energy and entropy when the $q\bar{q}$ pair is oriented perpendicular to the magnetic field. In this case, the string is parameterized in the static gauge $\tau=t$ as $\sigma=x_2$. This gives us
\begin{eqnarray}
\mathcal{F}^{\perp}(T,\ell) = \frac{L^2}{2 \pi \ell_{s}^2} \int\, dz \frac{e^{2A_s(z)}}{z^2} \sqrt{1+ \frac{g(z)q(z)}{z'^2}}\,,
\end{eqnarray}
where a prime $'$ now denotes a derivative with respect to $x_2$ and $q(z)=e^{B^2 z^2}$. There can again be connected and disconnected world sheet configurations that minimize the free energy. For the connected case, the free energy expression reduces to
\begin{equation}
\mathcal{F}_{con}^{\perp}=\frac{L^{2}}{ \pi l_{s}^{2}}\int_{0}^{z_{*}^{\perp}}  dz \, \left(\frac{z_{*}^{\perp}}{z^{2}}\right)^2 \frac{\sqrt{g(z)q(z)}e^{2A_{s}(z)-2A_{s}(z_{*}^{\perp})}}{\sqrt{g(z) q(z) z_{*}^{\perp 4}e^{-4A_{s}(z_{*}^{\perp})}-g(z_{*}^{\perp}) q(z_{*}^{\perp}) z^{4}e^{-4 A_{s}(z)}}} \,,
\label{freeEnergyconnperp}
\end{equation}
where the turning point of the connected string $z_{*}^{\perp}$ is related to the separation length $\ell^{\perp}$ as
\begin{equation}
\ell^{\perp}=2\int_{0}^{z_{*}^{\perp}}\,dz \sqrt{\frac{g(z_{*}^{\perp})q(z_{*}^{\perp})}{g(z)q(z)}}\frac{z^{2}e^{-2A_{s}(z)}}{\sqrt{g(z)q(z)z_{*}^{\perp 4}e^{-4A_{s}(z_{*}^{\perp})}-g(z_{*}^{\perp})q(z_{*}^{\perp}) z^{4}e^{-4 A_{s}(z)}}} \,.
\label{lengthperp}
\end{equation}
Similarly, the free energy expression of the disconnected string reduces to
\begin{equation}
\mathcal{F}_{discon}^{\perp}=\frac{L^{2}}{\pi l_{s}^{2}} \int_{0}^{z_h}\, dz\frac{e^{2A_{s}(z)}}{z^{2}}\,.
\end{equation}
Notice that the expression of $\mathcal{F}_{discon}^{\perp}$ is not only independent of $z_{*}^{\perp}$, and therefore of $\ell^{\perp}$, but also matches exactly with the $\mathcal{F}_{discon}^{\parallel}$ expression. This implies that the large distance behaviour of the $q\bar{q}$ free energy and entropy in a transverse magnetic field will remain the same as in the parallel case. Again, both $\mathcal{F}_{con}^{\perp}$ and $\mathcal{F}_{discon}^{\perp}$ contain UV divergences of the same nature, that cancelled out in the free energy difference $\Delta\mathcal{F}^{\perp}=\mathcal{F}_{con}^{\perp}-\mathcal{F}_{discon}^{\perp}$. In order to be consistent, we again regularize the $q\bar{q}$ free energies in the perpendicular case using the temperature-independent minimal renormalization scheme of \cite{Ewerz:2016zsx}.

\begin{figure}[h!]
\begin{minipage}[b]{0.5\linewidth}
\centering
\includegraphics[width=2.8in,height=2.3in]{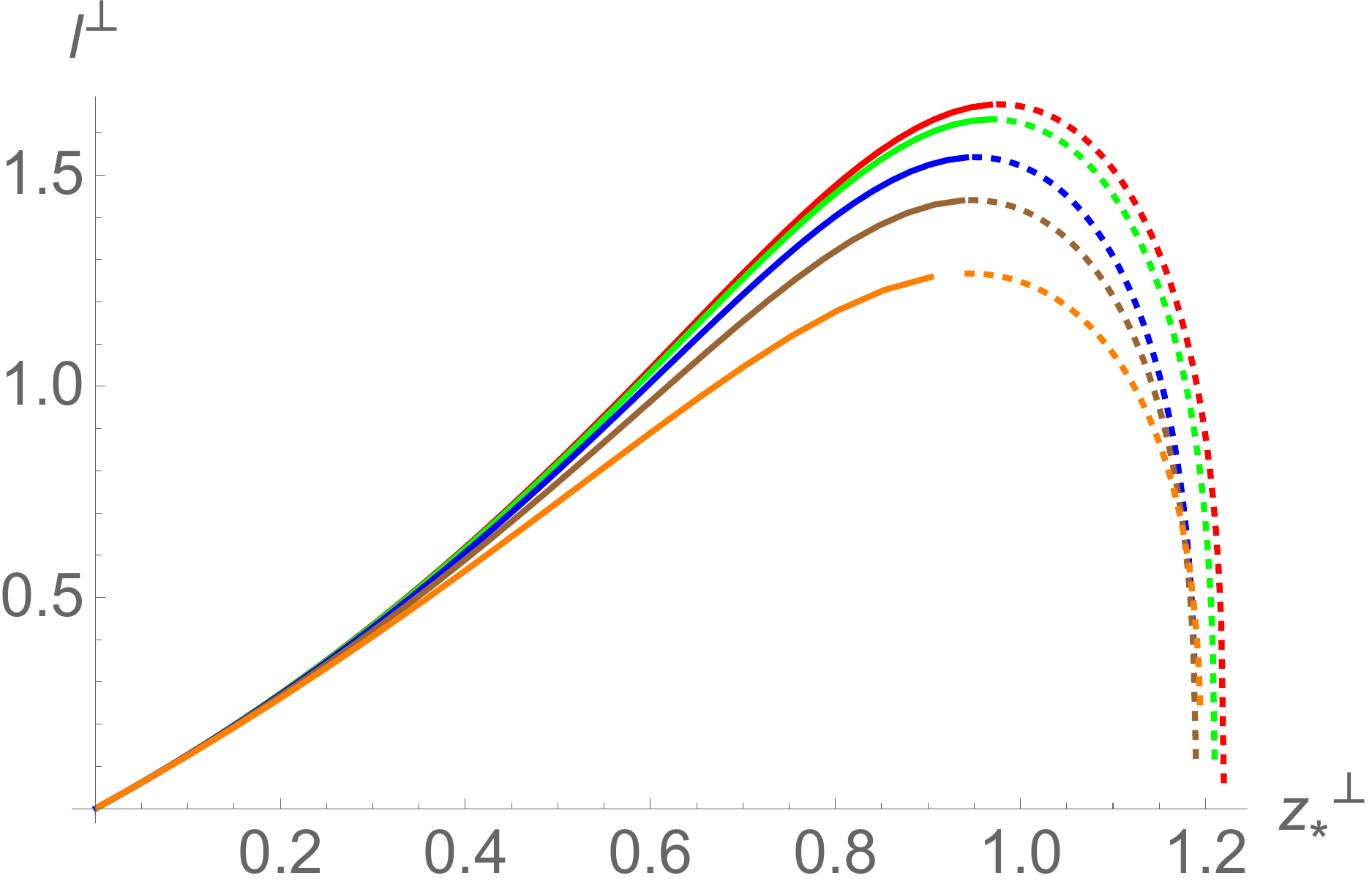}
\caption{ \small $\ell^{\perp}$ as a function of $z_{*}^{\perp}$ for various values of magnetic field. Here $T=1.2~T_{crit}$ is used. Red, green, blue, brown, and orange
curves correspond to $B=0$, $0.2$, $0.4$, $0.6$, and $0.8$ respectively. In units of GeV.}
\label{zsvslvsBT1Pt2TcperpendicularforA1}
\end{minipage}
\hspace{0.4cm}
\begin{minipage}[b]{0.5\linewidth}
\centering
\includegraphics[width=2.8in,height=2.5in]{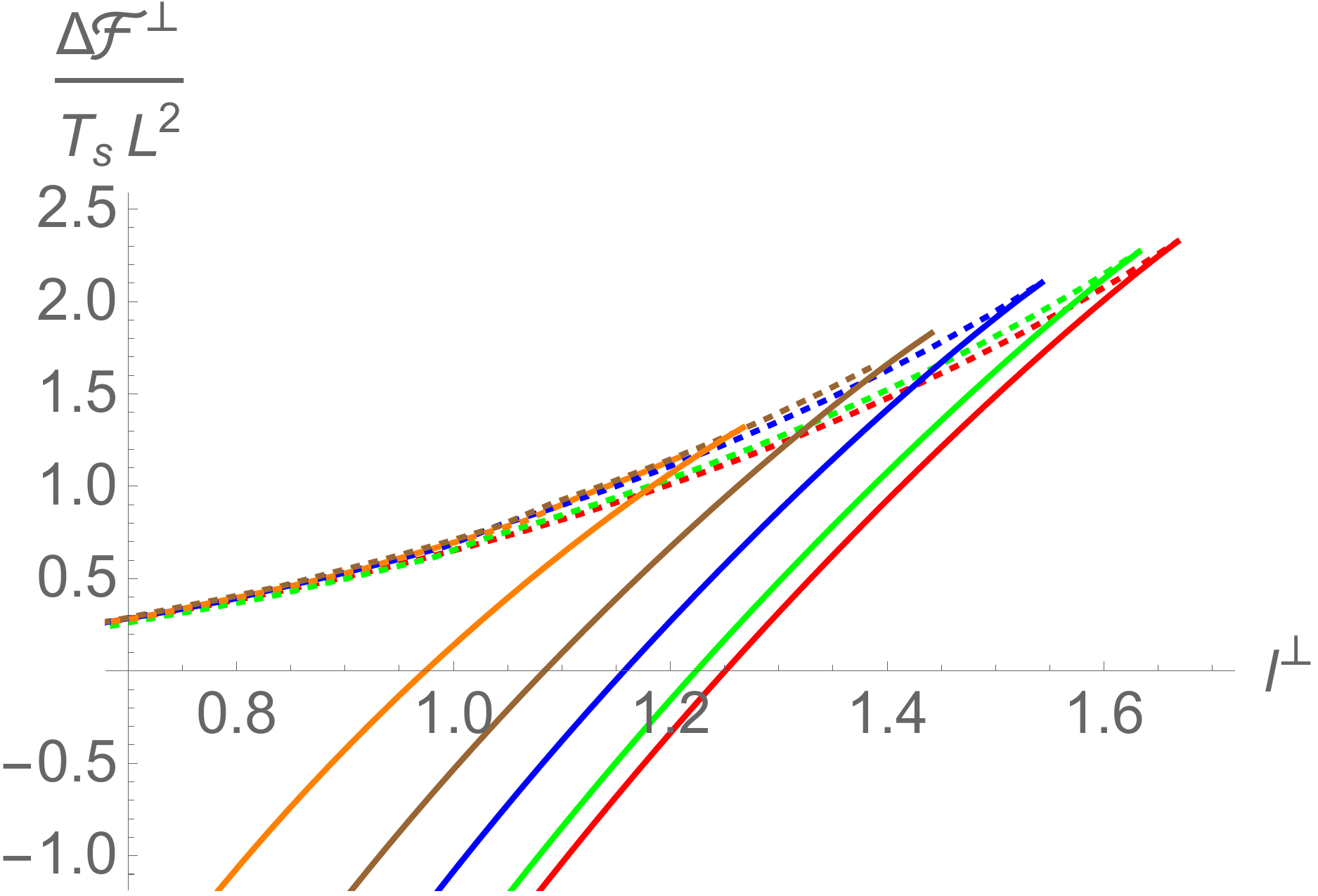}
\caption{\small $\Delta\mathcal{F}^{\perp}$ as a function of $\ell^{\perp}$ for various values of magnetic field. Here $T=1.2~T_{crit}$ is used. Red, green, blue, brown, and orange
curves correspond to $B=0$, $0.2$, $0.4$, $0.6$, and $0.8$ respectively. In units of GeV.}
\label{lvsDeltaFvsBT1Pt2TcperpendicularforA1}
\end{minipage}
\end{figure}

\begin{figure}[h!]
\centering
\includegraphics[width=2.8in,height=2.3in]{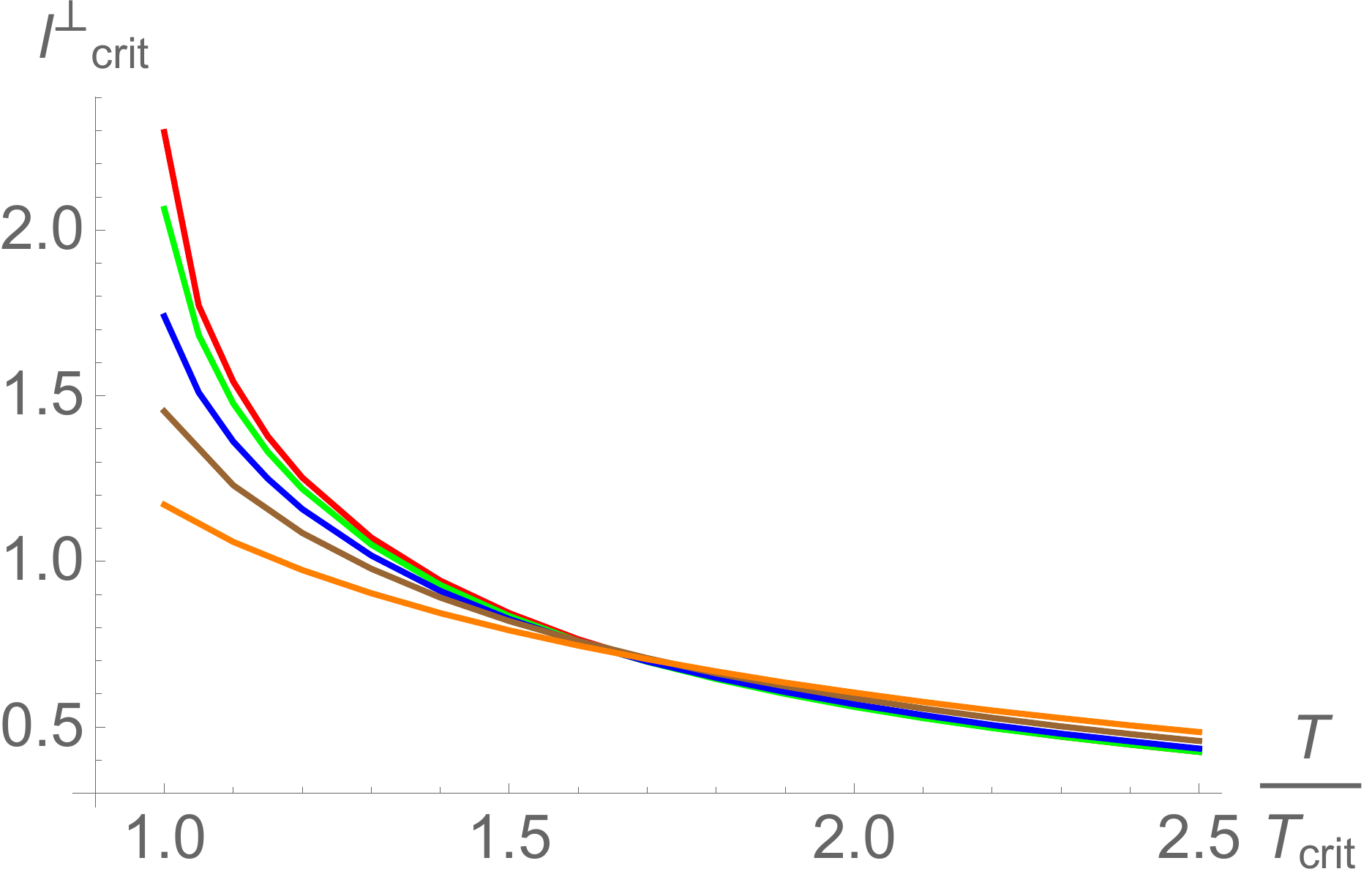}
\caption{ \small $\ell^{\perp}_{crit}$ as a function of temperature for various values of magnetic field. Red, green, blue, brown, and orange curves correspond to $B=0$, $0.2$, $0.4$, $0.6$, and $0.8$ respectively. In units of GeV.}
\label{TvslcritvsBperpendicularforA1}
\end{figure}

The numerical results for the perpendicular case are shown in Figures~\ref{zsvslvsBT1Pt2TcperpendicularforA1} and \ref{lvsDeltaFvsBT1Pt2TcperpendicularforA1}, where the $z_{*}^{\perp}$ vs. $\ell^{\perp}$ and $\ell^{\perp}$ vs. $\Delta\mathcal{F}^{\perp}$ behaviour are illustrated, respectively, for various magnetic field values. These results are similar to the parallel case discussed above. In particular, there are again two solutions of the turning point $z_{*}^{\perp}$ for a given $\ell^{\perp}< \ell^{\perp}_{max}$, and these two solutions cease to exist above $\ell^{\perp}_{max}$. The magnitude of $\ell^{\perp}_{max}$ moreover decreases with the magnetic field for temperatures near the deconfinement temperature. Similarly, $\Delta\mathcal{F}^{\perp}$ undergoes a sign change as the separation length is varied, suggesting a phase transition between connected and disconnected free energies. The critical separation length $\ell^{\perp}_{crit}$ at which this phase transition occurs again depends nontrivially on the magnetic field and temperature. This is shown in Figure~\ref{TvslcritvsBperpendicularforA1}, where the complete dependence of $\ell^{\perp}_{crit}$ on $T$ and $B$ is shown. We moreover find that $\ell_{crit}$ for the parallel case is slightly higher than the transverse case at a fixed temperature, though the difference is negligible for small magnetic field values. Similarly, the difference between $\ell_{crit}^{\parallel}$ and $\ell_{crit}^{\perp}$ is appreciable only near the deconfinement temperature whereas away from it the difference becomes negligible.

\begin{figure}[h!]
\begin{minipage}[b]{0.5\linewidth}
\centering
\includegraphics[width=2.8in,height=2.3in]{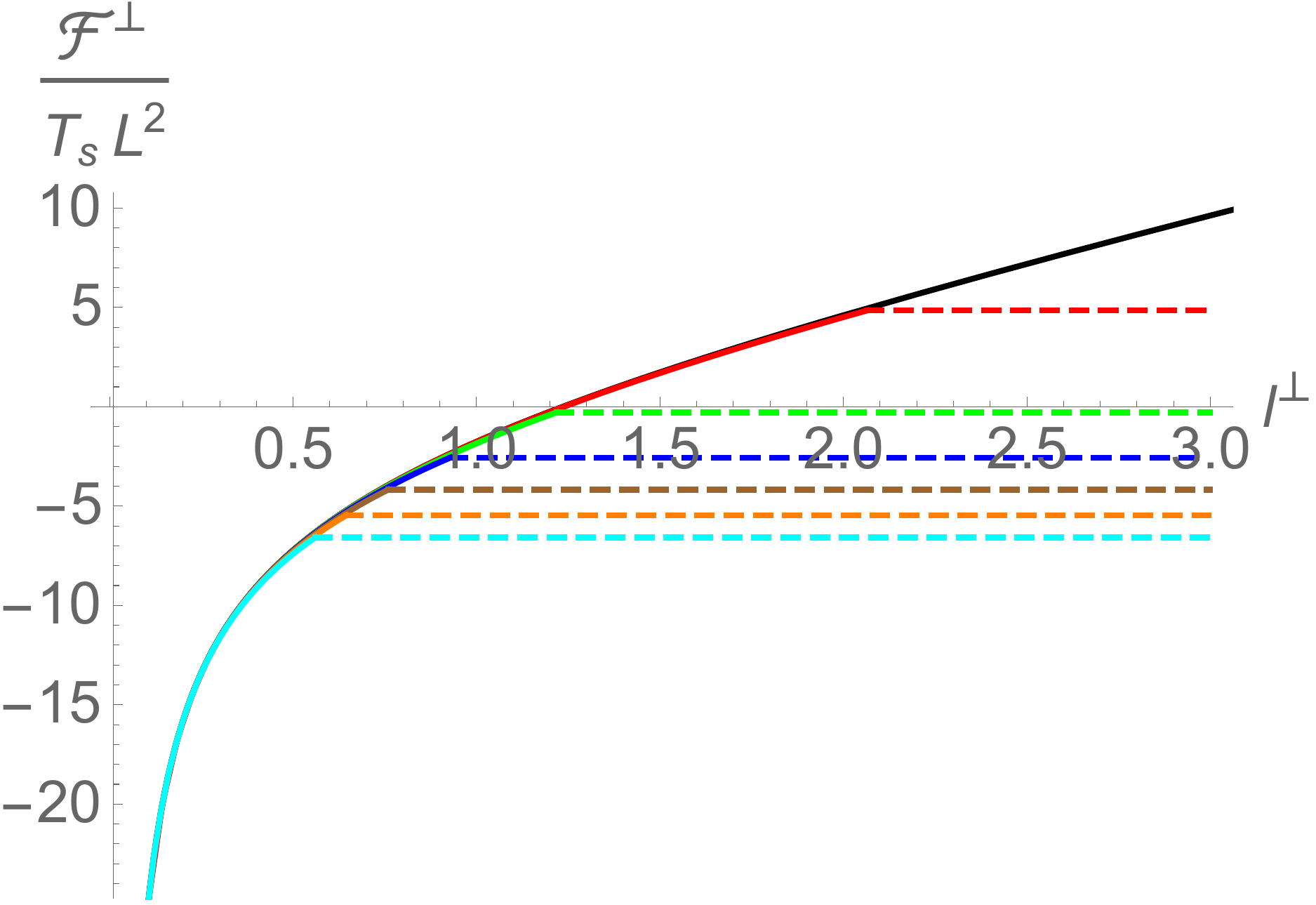}
\caption{ \small $\mathcal{F}^{\perp}$ as a function of $\ell^{\perp}$ for various values of temperature. Here $B=0.2$ is used. Black, red, green, blue, brown, orange, and cyan
curves correspond to $T/T_{crit}=0$, $1.0$, $1.2$, $1.4$, $1.6$, $1.8$, and $2.0$ respectively. In units of GeV.}
\label{lvsFvsTBPt2perpendicularforA1}
\end{minipage}
\hspace{0.4cm}
\begin{minipage}[b]{0.5\linewidth}
\centering
\includegraphics[width=2.8in,height=2.5in]{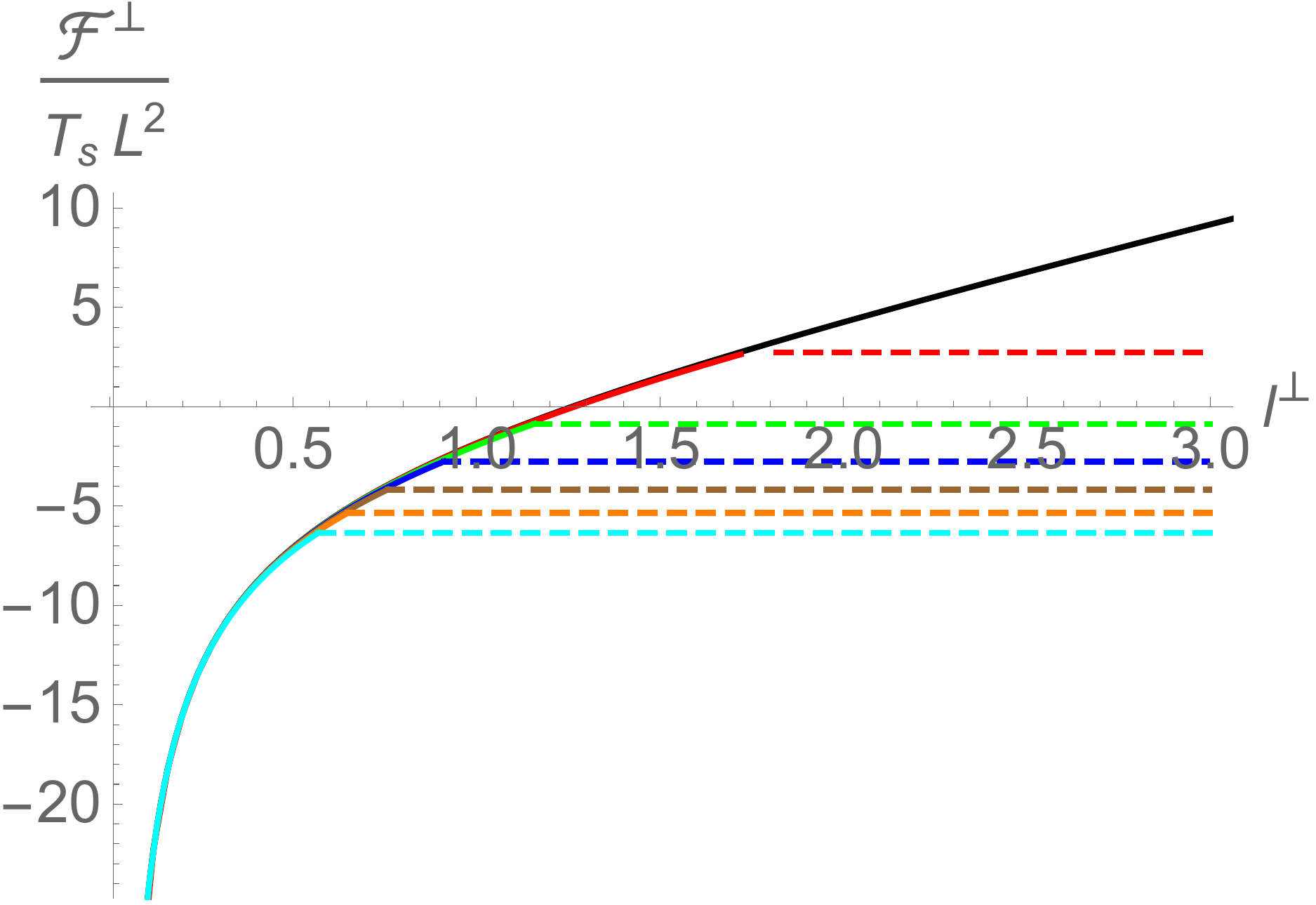}
\caption{\small $\mathcal{F}^{\perp}$ as a function of $\ell^{\perp}$ for various values of temperature. Here $B=0.4$ is used. Black, red, green, blue, brown, orange, and cyan
curves correspond to $T/T_{crit}=0$, $1.0$, $1.2$, $1.4$, $1.6$, $1.8$, and $2.0$ respectively. In units of GeV.}
\label{lvsFvsTBPt4perpendicularforA1}
\end{minipage}
\end{figure}

We further analyse the thermal profile of $q\bar{q}$ free energy in the transverse case. This is shown in Figures~\ref{lvsFvsTBPt2perpendicularforA1} and \ref{lvsFvsTBPt4perpendicularforA1}. We again find a Coulombic structure in the $q\bar{q}$ free energy at small $\ell^{\perp}$ for all $T$ and $B$. As the separation length increases, the free energy also increases and then saturates to a constant value at large separations for all $T$ and $B$. Interestingly, the saturation value of the free energy at large separation again decreases with $B$ for temperatures near the deconfinement temperature whereas it increases with $B$ at much higher temperatures. The free energy further decreases with temperature for a fixed $\ell^{\perp}$ and $B$, implying again thermal screening of color charges.

We can similarly analyse the entropy of the $q\bar{q}$ pair in the perpendicular case. Like in the parallel case, we again have different expressions of the $q\bar{q}$ entropy at small and large separations. In particular, for small separations, we have
\begin{equation}
S^{\perp}_{con}(\ell^{\perp}<\ell^{\perp}_{crit})=-\frac{\partial\mathcal{F}^{\perp}_{con}}{\partial T}\,,
\label{Sperpendicularcon}
\end{equation}
whereas, for large separations, we have
\begin{equation}
S^{\perp}_{discon}(\ell^{\perp}>\ell^{\perp}_{crit})=-\frac{\partial\mathcal{F}^{\perp}_{discon}}{\partial T}\,.
\label{Sperpendiculardiscon}
\end{equation}

The thermal profile of the $q\bar{q}$ entropy at large separations in the perpendicular case is exactly similar to the parallel case (see Figure~\ref{TvsSvsBlargeLparallelforA1}). This is due to the fact that at large separations the entropy is given by the disconnected free energy, which expression matches in both cases. This further implies that in the perpendicular case as well, the $q\bar{q}$ pair exhibits a peak and a large amount of entropy at the deconfinement temperature. Accordingly, other interesting results such as a substantial decrement of the entropy at the peak in the presence of magnetic field (compared to the zero magnetic field case)  or higher temperature asymptotic structure of the entropy remain true in the perpendicular case.

\begin{figure}[h!]
\centering
\includegraphics[width=2.8in,height=2.3in]{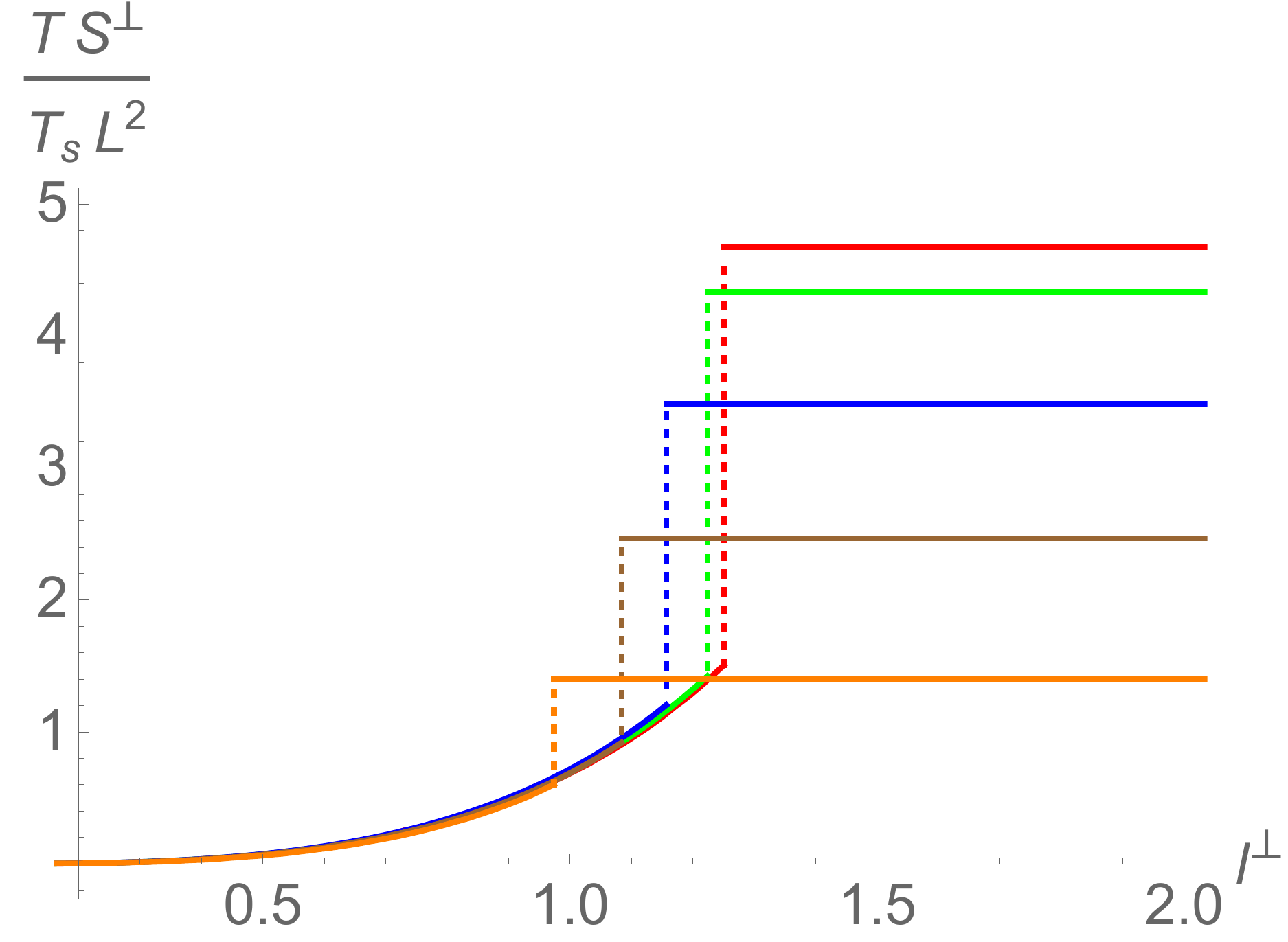}
\caption{ \small $S^{\perp}$ as a function of $\ell^{\perp}$ for various values of magnetic field. Here $T=1.2~T_{crit}$ is used. Red, green, blue, brown, and orange curves correspond to $B=0$, $0.2$, $0.4$, $0.6$, and $0.8$ respectively. In units of GeV.}
\label{lvsSvsBT1Pt2TcperpendicularforA1}
\end{figure}

The entropy profile with separation length, as shown in Figure~\ref{lvsSvsBT1Pt2TcperpendicularforA1}, similarly exhibits analogous features in the perpendicular case. We find that $T S^{\perp}$ first increases with $\ell^{\perp}$, and then saturates to a constant value at large $\ell^{\perp}$ in a discontinuous manner. This is again due to the fact that the free energy of the disconnected string, which is the relevant string configuration, is independent of $\ell^{\perp}$. Moreover, like in the parallel case, the saturation value of the entropy decreases with the magnetic field. This is true for high temperatures as well. The main difference compared to the parallel case arises at the length scale at which the entropy saturates. This is given by $\ell_{crit}$. As mentioned before, though the difference between $\ell_{crit}^{\parallel}-\ell_{crit}^{\perp}$ is small and positive, however, this difference can be appreciable at large $B$. This suggests that for parallel case the entropy  saturates at larger lengths compared to the perpendicular case for large magnetic field values.

\begin{figure}[h!]
	\centering
	\includegraphics[width=2.8in,height=2.3in]{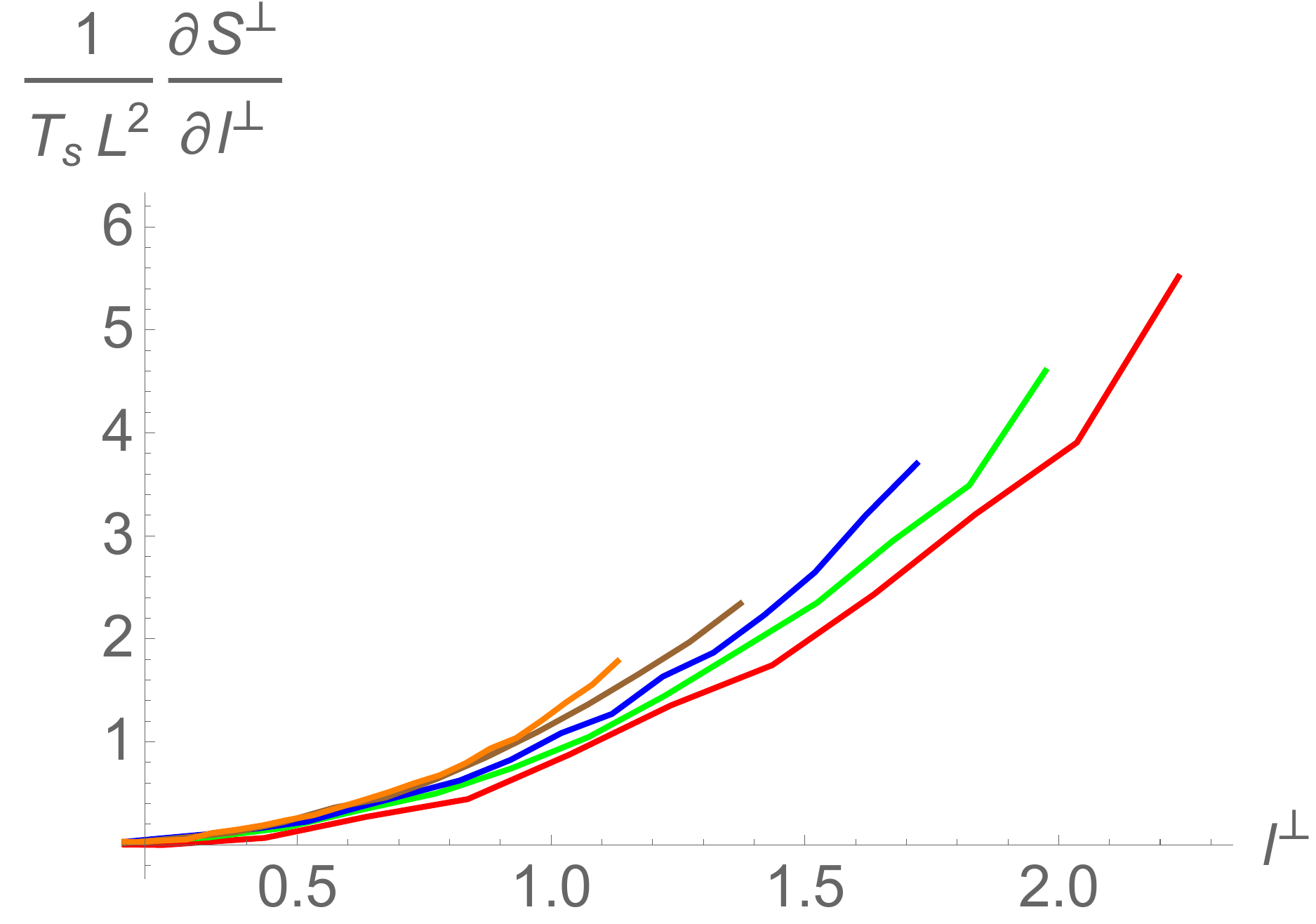}
	\caption{ \small Entropic force  as a function of $\ell^{\perp}$ for various values of magnetic field in the perpendicular case. Here $T=1.0~T_{crit}$ is used. Red, green, blue, brown, and orange curves correspond to $B=0$, $0.2$, $0.4$, $0.6$, and $0.8$ respectively. In units of GeV.}
	\label{lvsEntropicforcevsBT1TcperpendicularforA1}
\end{figure}

We further analyse the slope of $\ell^{\perp}$ vs.~$S^{\perp}$ near the deconfinement temperature and find it to be increasing with the magnetic field in the perpendicular case as well. This is illustrated in Figure~\ref{lvsEntropicforcevsBT1TcperpendicularforA1}. This analysis again indicates an enhancement of the entropic force (thereby suggesting a higher dissociation of the heavy quark pair in the presence of a magnetic field) and once again signal towards the inverse magnetic response of the system in the dissociation sector.

Before ending this section we would like to mention that our analysis suggests limited anisotropic effects of the magnetic field on the high-temperature deconfined phase compared to the low temperature confined phase. In particular, though some non-trivial changes do arise between parallel and transverse $q\bar{q}$ free energies (and entropies) at above-deconfined temperatures, however, these changes are not as substantial as in the $T=0$ confined phase. For instance, the magnetic field creates distinct and measurable effects in the string tension of the confined phase in parallel and transverse directions, i.e., the string tension decreases/increases in parallel/transverse directions \cite{Bohra:2019ebj,Bonati:2014ksa}. In the deconfinement phase, the anisotropic effects due to the magnetic field might be getting suppressed by the large thermal effects.  One can also do a large temperature expansion and find that the effect of the magnetic field only appears at higher orders for large temperatures. In any case, our holographic results for the $q\bar{q}$ free energy and entropy in the deconfined phase at $B=0$ do match qualitatively well with lattice results. Interestingly, our analysis further predicts the occurrence of similar features in the $q\bar{q}$ free energy and entropy in the presence of a magnetic field, irrespective of its orientation, as well. As we will see in the next section, the rather limited effects of the orientation of the magnetic field in the deconfined phase continue to hold for the quarkonium dissociation time scale as well.

\section{Real-time dynamics of quarkonium dissociation}
We now discuss the physically appealing question of the real-time dynamics of quarkonium dissociation, a quantity that is not directly accessible via lattice QCD. As mentioned earlier, the holographic prescription of quarkonium dissociation is given by a string with one end fixed at the AdS boundary and the other end at the black hole horizon. When the string reaches the horizon, the system reaches its thermal equilibrium state in which the string is split into two pieces. The time needed for the string to reach the horizon will give us an approximate time scale for the quarkonium dissociation \cite{Iatrakis:2015sua} \footnote{To be more precise, the time needed for the string to reach the horizon will only give an upper bound on the dissociation time. For the correct and more refined dissociation time, one actually needs to find the time when the connected configuration breaks into the disconnected configuration when the string falls due to the black hole gravitational pull. However, this analysis is numerically more challenging as it would require delicate care of the spatial variation part of the string equation of motion. Here we choose to neglect the spatial variation part for simplicity and continue to call the time needed for the string to reach the horizon as the dissociation time. The analogous analysis including the spatial variation part will be discussed in future work.}. To compute this, we must once again consider a string of length $\ell$ connecting the $q\bar{q}$ pair at the boundary. We again have a choice to appropriately parameterize the string depending upon whether it is oriented parallel or perpendicular to the magnetic field. For the parallel case, we consider the parametrization $X^M=(t,z(t,x_1),x_1,0,0)$. The Nambu-Goto action then reads
\begin{eqnarray}
S^{\parallel}_{NG} = \frac{L^2}{2 \pi \ell_{s}^2} \int\, dt dx_1 \ \frac{e^{2A_s(z)}}{z^2} \sqrt{g(z)+ z'^2-\frac{\dot{z}^2}{g(z)}}\,,
\label{NGactionparallelRealtime}
\end{eqnarray}
where $\dot{z}=\partial_t z$, and $z'=\partial_{x_1} z$. The equation of motion of the string then reads,
\begin{eqnarray}
 -\partial_t \left(\frac{e^{2A_s(z)}}{z^2 g(z)} \frac{\dot{z}}{\sqrt{g(z)+ z'^2-\frac{\dot{z}^2}{g(z)}}} \right) +\partial_{x_1} \left(\frac{e^{2A_s(z)}}{z^2} \frac{z'}{\sqrt{g(z)+ z'^2-\frac{\dot{z}^2}{g(z)}}}  \right) \nonumber \\
 -\partial_z \left(\frac{e^{2A_s(z)}}{z^2} \right) \sqrt{g(z)+ z'^2-\frac{\dot{z}^2}{g(z)}} - \frac{e^{2A_s(z)}\partial_z g(z)}{2 z^2 \sqrt{g(z)+ z'^2-\frac{\dot{z}^2}{g(z)}}}\left(1+\frac{\dot{z}^2}{g(z)^2}   \right) = 0\,.
\label{StringEOMparallelRealtime}
\end{eqnarray}
We can numerically solve the above equation with suitable boundary conditions. For this purpose, we consider the fact that the end points of the strings are fixed on the boundary $z(t,x_1=\pm\ell^{\parallel}/2)=\varepsilon_{UV}$, and that the string is initially on the boundary  $z(t=0,x_1)=\varepsilon_{UV}$ at rest, $\dot{z}(t=0,x_1)=\varepsilon_{UV}$. Here $\varepsilon_{UV}$ as usual is the boundary cut-off. In principle, with these boundary conditions, Eq.~(\ref{StringEOMparallelRealtime}) can be solved for all $\ell^{\parallel}$. However, for small $\ell^{\parallel}$, the spatial variation part gives a large numerical error and makes this partial differential equation difficult to solve.  The difficulty at small $\ell^{\parallel}$ is further amplified by the fact that our coordinate system is singular at the black hole horizon and, therefore, not really appropriate to study dynamics near the horizon. Accordingly, utmost numerical care is required while discussing dynamics for small $\ell^{\parallel}$.  On the other hand, for large $\ell^{\parallel}$, as the spatial variation is expected to be negligible (remember that for large, the $\ell^{\parallel}$ string profile is given by the disconnected configuration which is independent of $x_1$), above mentioned numerical subtleties do not arise. In this case, we can safely neglect the spatial variation part in the Nambu-Goto action [Eq.~(\ref{NGactionparallelRealtime})]. This provides a large simplification in the string real-time dynamics, see also \cite{Iatrakis:2015sua}. Moreover, for the quarkonium dissociation (which is mainly described by the disconnected string configuration in holographic context), it is indeed fairly reasonable to discuss dynamics using only large $\ell^{\parallel}$ physics. For these reasons, here we will mainly concentrate on the string dynamics with large interquark separation. In this simpler case, the string action takes the form $S^{\parallel}_{NG} = \frac{L^2}{2 \pi \ell_{s}^2} \int\, dt dx_1 \ \frac{e^{2A_s(z)}}{z^2} \sqrt{g(z)-\frac{\dot{z}^2}{g(z)}}$. A similar simpler analysis was also adopted in \cite{Iatrakis:2015sua}.  The corresponding dynamics with spatial variation for small interquark separation will be addressed in a separate work.

The string equation of motion with neglected spatial variation reads,
\begin{eqnarray}
& & -\partial_t \left(\frac{e^{2A_s(z)}}{z^2 g(z)} \frac{\dot{z}}{\sqrt{g(z)-\frac{\dot{z}^2}{g(z)}}} \right) -\partial_z \left(\frac{e^{2A_s(z)}}{z^2} \right) \sqrt{g(z)-\frac{\dot{z}^2}{g(z)}}  \nonumber \\
& &  - \frac{e^{2A_s(z)}\partial_z g(z)}{2 z^2 \sqrt{g(z)-\frac{\dot{z}^2}{g(z)}}}\left(1+\frac{\dot{z}^2}{g(z)^2}  \right) = 0\,.
\label{StringEOMparallelRealtime1}
\end{eqnarray}
Since the action does not explicitly depend on time, there exists a corresponding conserved energy,
\begin{eqnarray}
 E^{\parallel}  = T_s \frac{e^{2A_s(z)}}{z^2} \frac{g(z)}{\sqrt{g(z)-\frac{\dot{z}^2}{g(z)}}}\,.
\label{StringE}
\end{eqnarray}
In terms of $E^{\parallel}$, the velocity of the string can be expressed as
\begin{equation}
\dot{z}= \frac{g(z)}{E}\sqrt{(E^{\parallel})^{2}- \frac{T_{s}^2 e^{4 A_{s}(z)}g(z)}{z^4}}
\end{equation}
We can moreover fix $E^{\parallel}$ from the boundary condition that string falls from the boundary with zero initial velocity $\dot{z}(t=0)=0$. Putting the boundary cut-off at $z=\varepsilon_{UV}$, the energy of the falling string can be expressed as
\begin{equation}
E^{\parallel} = \frac{T_{s} e^{2 A_{s}(\varepsilon_{UV})}\sqrt{g(\varepsilon_{UV})}}{\varepsilon_{UV}^2}\,.
\end{equation}

\begin{figure}[h!]
\centering
\includegraphics[width=2.8in,height=2.3in]{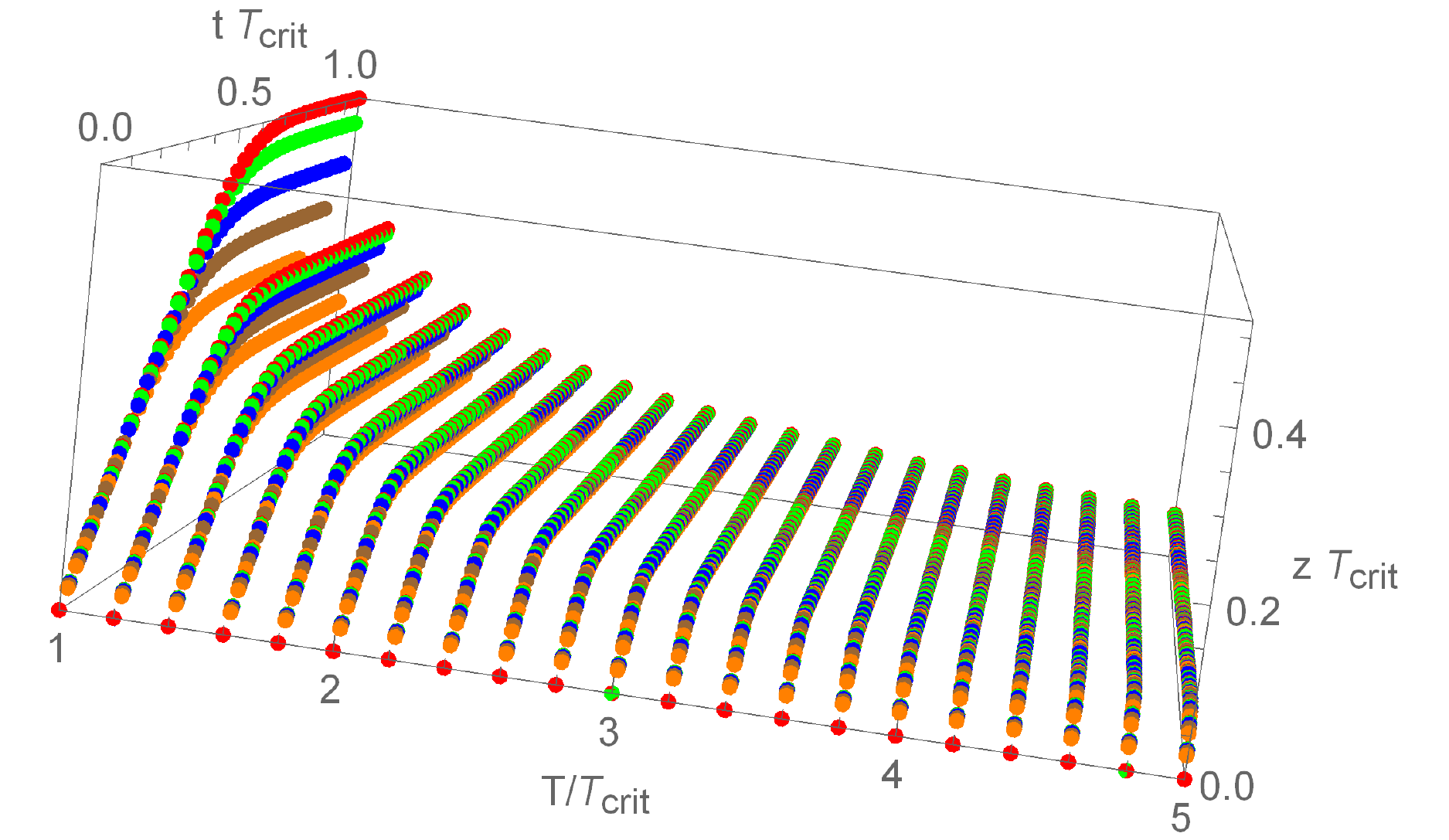}
\caption{ \small The string motion from the boundary to the horizon for various values of parallel magnetic field and temperature. Red, green, blue, brown, and orange curves correspond to $B=0$, $0.2$, $0.4$, $0.6$, and $0.8$ respectively. In units of GeV.}
\label{tvsTvsBvsZprofileparallelforA1}
\end{figure}

We can now find the string's entire motion by integrating its equation of motion from the boundary to the horizon numerically. The numerical output is shown in Figure~\ref{tvsTvsBvsZprofileparallelforA1}, where the string position as a function of time for different magnetic field and temperature values is plotted. Notice that for all cases the string accelerates faster near the boundary and then asymptotically approaches the horizon. With the magnetic field, substantial changes in string dynamics appear only near the deconfinement temperature.  For higher temperatures, the string profiles for different magnetic fields overlap with each other and they approach the horizon at a faster rate. This is expected from the fact that higher temperature corresponds to larger black holes having a stronger gravitational pull. In fact, using the near horizon behaviour of various fields, it can be shown that the string approaches the horizon exponentially fast,
\begin{equation}
z(t)-z_h\simeq e^{-4\pi T t}\,.
\label{nearhorizonapproach}
\end{equation}
Since for a fixed $z_h$ the temperature decreases with the magnetic field, the above equation also indicates that the string approaches the fixed size horizon at a slower rate as the magnetic field increases.

We now compute the time required for the dissociation of heavy quarkonium. This is approximated as the time required for the string to reach the horizon from the asymptotic boundary \cite{Iatrakis:2015sua}. A closed expression of the dissociation time $t^{\parallel}_D$ can be found from the string equation of motion, giving
\begin{equation}
t^{\parallel}_{D}= \int_{z=\varepsilon_{UV}}^{z_h} dz \frac{1}{g(z)\sqrt{1- \frac{T_{s}^2 e^{4 A_{s}(z)}g(z)}{(E^{\parallel})^2 z^4}}}\,.
\end{equation}
The upper limit of this integral however makes $t^{\parallel}_D$ infinite (because of the factor $g(z)$ in the denominator). This is expected since the string is approaching the horizon asymptotically. Therefore, to get a sensible result, we put an IR cut-off just outside the horizon $\varepsilon_{IR}=10^{-4}$, i.e., we replace the upper limit of the above integral by $z_h=z_h-\varepsilon_{IR}$. This can be recognized as the point where the string thermalizes. Therefore, by varying $z_h$ and $B$, the temperature and magnetic field dependence of the dissociation time can be studied.

\begin{figure}[h!]
\centering
\includegraphics[width=2.8in,height=2.3in]{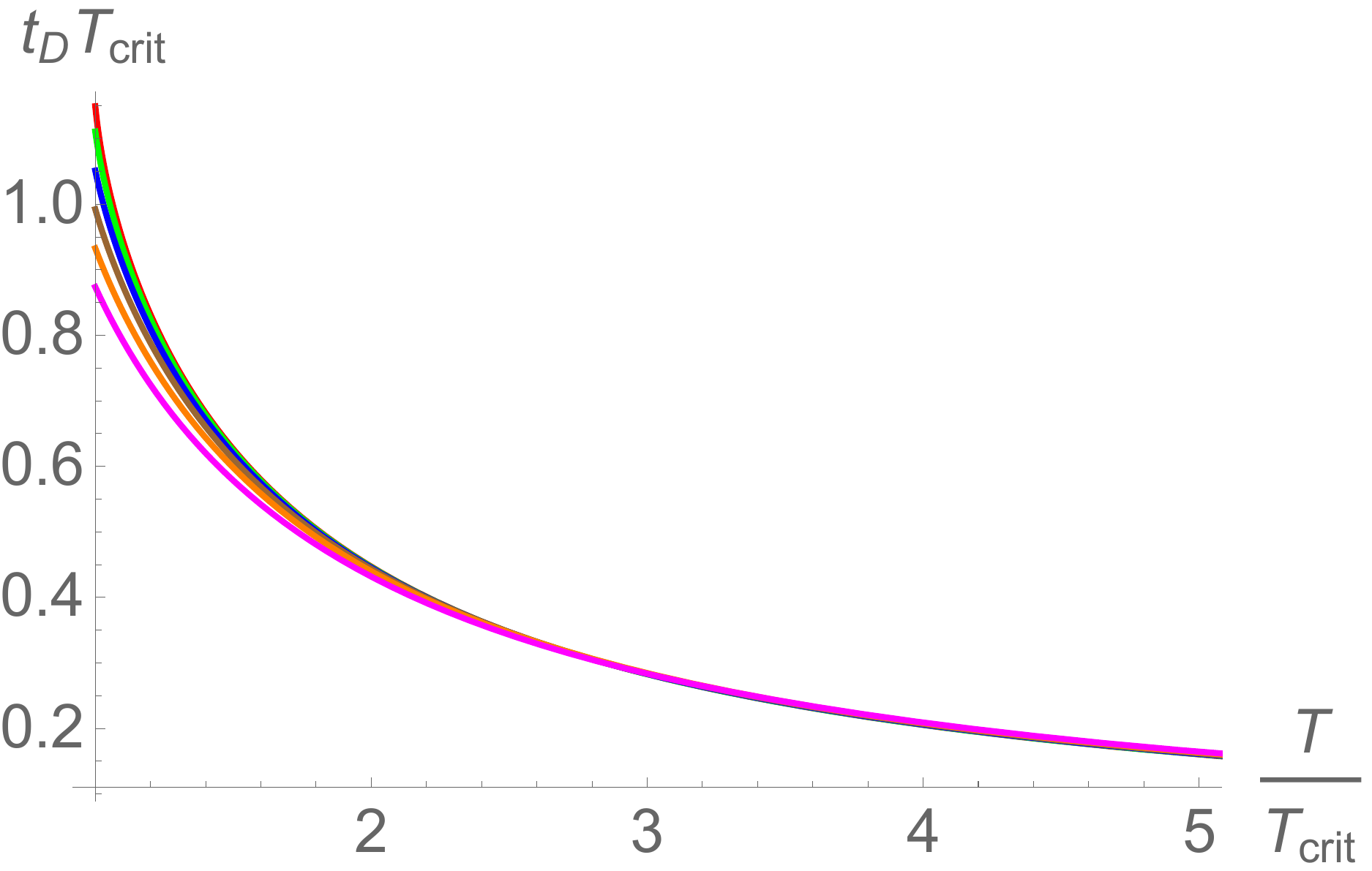}
\caption{ \small The quarkonium dissociation time as a function of temperature for various values of parallel magnetic field. Red, green, blue, brown, orange, and magenta curves correspond to $B=0$, $0.2$, $0.4$, $0.6$, $0.8$, and $1.0$ respectively.}
\label{TvstDvsBparallelforA1}
\end{figure}

In Figure~\ref{TvstDvsBparallelforA1}, the thermal profile of $t^{\parallel}_D$ for different magnetic field values is plotted \footnote{Here the thermal and magnetic field dependence of the dissociation time is plotted with rescaled units (in a dimensionless manner) for simplicity. This result can also be easily generalised to GeV units, i.e.~without rescaling of the axes. Needless to say that for both (with and without rescaling), the dissociation time at a fixed temperature decreases with the magnetic field.}. We observe that the dissociation time at a fixed magnetic field is shorter for higher temperatures. This is physically expected as the string would fall faster towards the horizon for larger black holes. Importantly, our analysis further suggests a decrement in the dissociation time as the magnetic field is turned on. In particular, it indicates a faster dissociation of heavy quarkonium as the magnetic field is increased. This result can again be taken as an indication of inverse magnetic catalysis, but now at the dynamical level. This result could further be considered as an extrapolation from the confined phase results where the string tension was found to be decreasing with the parallel magnetic field. Moreover, notice that the dissociation time around the deconfinement temperature is less than a fermi, suggesting an extremely fast dissociation process of heavy quarkonium around that temperature.

\begin{figure}[h!]
\centering
\includegraphics[width=2.8in,height=2.3in]{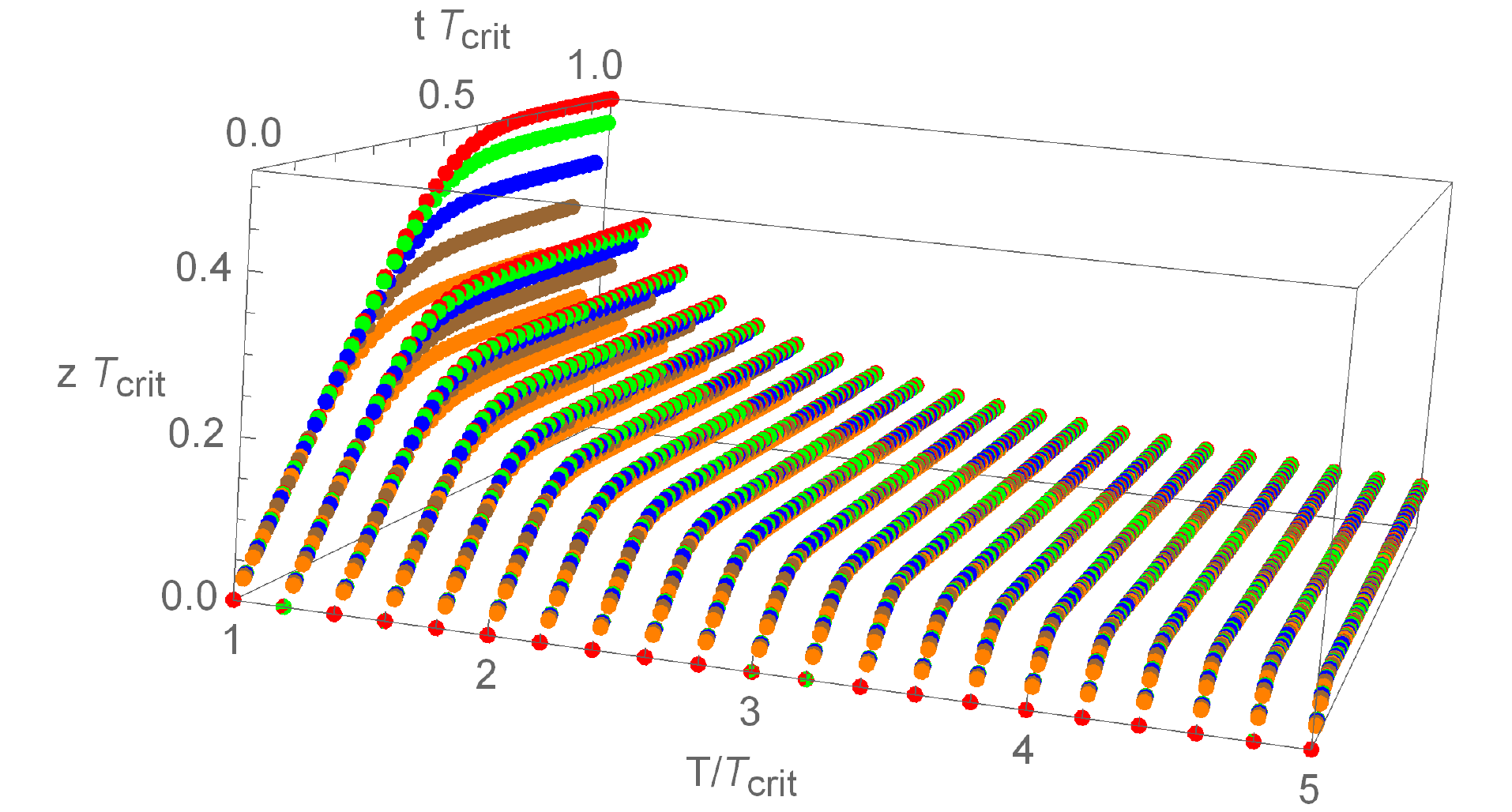}
\caption{ \small The string motion from the boundary to the horizon for various values of transverse magnetic field and temperature. Red, green, blue, brown, and orange curves correspond to $B=0$, $0.2$, $0.4$, $0.6$, and $0.8$ respectively. In units of GeV.}
\label{tvsTvsBvsZprofileperpendicularforA1}
\end{figure}

We can similarly discuss the real-time dynamics of the string with a transverse magnetic field. For the transverse parametrization $X^M=(t,z(t,x_2),0,x_2,0)$,  the Nambu-Goto action for large interquark separation reads
\begin{eqnarray}
S^{\perp}_{NG} = \frac{L^2}{2 \pi \ell_{s}^2}\int dt dx_2 \frac{e^{2 A_{s}(z)}e^{\frac{B^2 z^2}{2}}}{z^2} \sqrt{g(z)-\frac{\dot{z}^2}{g(z)}}\,.
\label{StringEOMparallelRealtime1}
\end{eqnarray}
This gives the following equation of motion for the string,
\begin{eqnarray}
 -\partial_t \left(\frac{e^{2A_s(z)+ \frac{B^2 z^2}{2}}}{z^2 g(z)} \frac{\dot{z}}{\sqrt{g(z)-\frac{\dot{z}^2}{g(z)}}} \right) -\partial_z \left(\frac{e^{2A_s(z)+\frac{B^2 z^2}{2}}}{z^2} \right) \sqrt{g(z)-\frac{\dot{z}^2}{g(z)}}  \nonumber \\
  - \frac{e^{2A_s(z)+\frac{B^2 z^2}{2}}\partial_z g(z)}{2 z^2 \sqrt{g(z)-\frac{\dot{z}^2}{g(z)}}}\left(1+\frac{\dot{z}^2}{g(z)^2}  \right) = 0\,,
\label{StringEOMperpendicularRealtime1}
\end{eqnarray}
having the conserved energy
\begin{eqnarray}
 E^{\perp}  = T_s \frac{e^{2A_s(z)+\frac{B^2 z^2}{2}}}{z^2} \frac{g(z)}{\sqrt{g(z)-\frac{\dot{z}^2}{g(z)}}}\,,
\label{StringE}
\end{eqnarray}
which could again be fixed from the boundary condition $\dot{z}(t=0)=0$. Using the same numerical strategy as illustrated for the parallel case we can study the string dynamics in the transverse case. The numerical results are shown in Figure~\ref{tvsTvsBvsZprofileperpendicularforA1}, where the string motion as a function of time is shown. For the transverse case as well the string accelerates quickly near the boundary and then asymptotically approaches the horizon. The approach to the horizon is again of an exponential nature as in Eq.~(\ref{nearhorizonapproach}).

Similarly, the expression for the dissociation time now reduces to
\begin{equation}
t^{\perp}_{D}= \int_{z=\varepsilon_{UV}}^{z_h-\varepsilon_{IR}} dz \frac{1}{g(z)\sqrt{1- \frac{T_{s}^2 e^{4 A_{s}(z)+B^2 z^2}g(z)}{(E^\perp)^2 z^4}}}\,.
\end{equation}

\begin{figure}[h!]
\centering
\includegraphics[width=2.8in,height=2.3in]{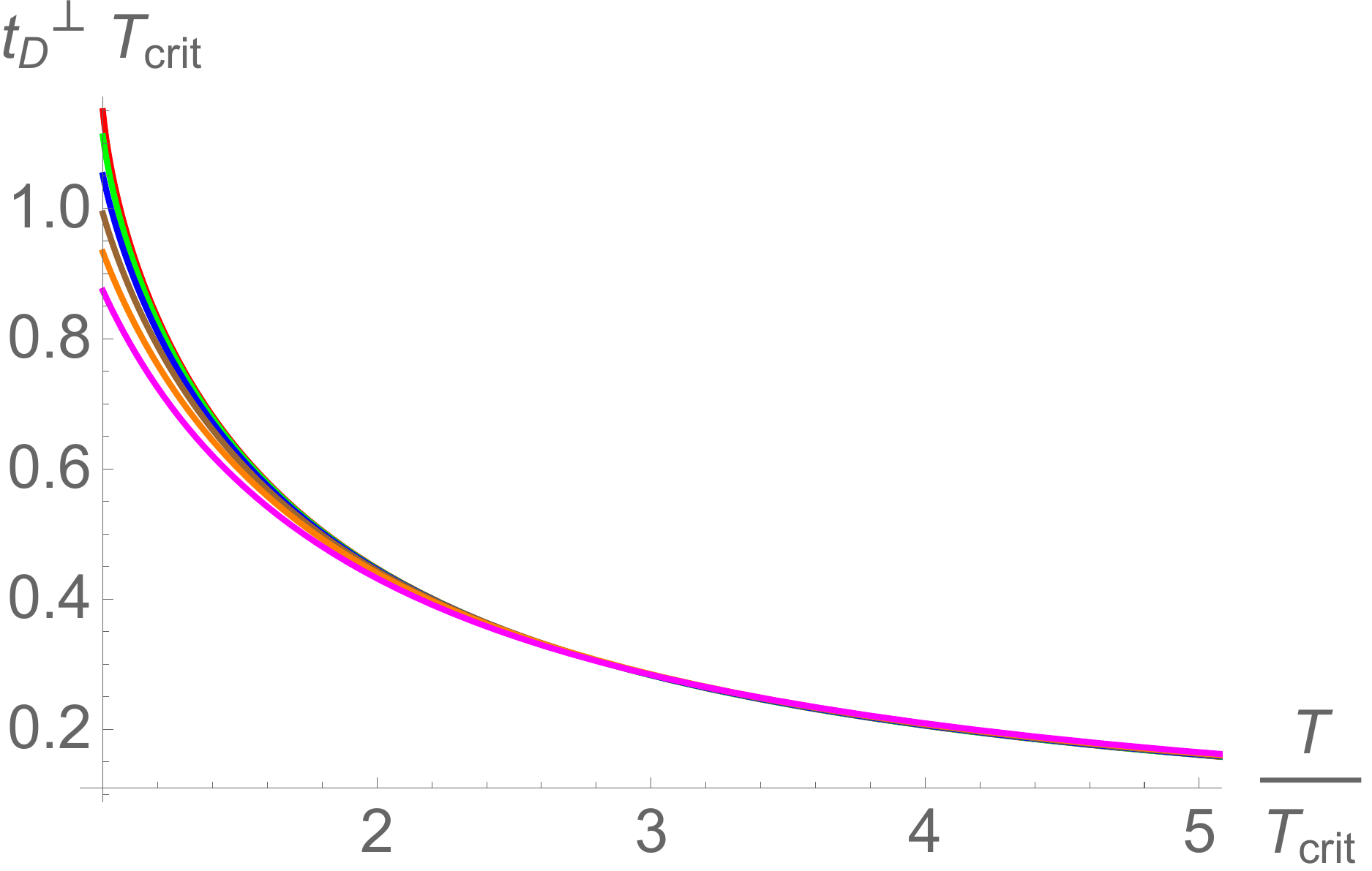}
\caption{ \small The quarkonium dissociation time as a function of temperature for various values of transverse magnetic field. Red, green, blue, brown, orange, and magenta curves correspond to $B=0$, $0.2$, $0.4$, $0.6$, $0.8$, and $1.0$ respectively.}
\label{TvstDvsBperpendicularforA1}
\end{figure}

The results for the  dissociation time for the transverse case are shown in Figure~\ref{TvstDvsBperpendicularforA1}. Most of the results remain the same as in the parallel case. In particular, the dissociation time decreases not only with temperature but also with magnetic field. Therefore, our analysis suggests inverse catalysis behaviour in the heavy quarkonium dissociation with a magnetic field, irrespective of its orientation. The difference between the parallel and transverse dissociation times is moreover found to be negligible, i.e.,  $t^{\perp}_{D} \simeq t^{\parallel}_{D}$. This again indicates a limited anisotropic effect of the magnetic field in the high-temperature deconfined phase. As we have seen thoroughly, though a magnetic field produces substantial changes in various QCD scales, like the screening length or the dissociation time, however, its orientation related effects are rather negligible in the high-temperature deconfined phase. This is indeed an important result that could be tested in lattice settings.

\section{Conclusions}
In this work, we investigated the effects of a magnetic field on the $q\bar{q}$ free energy, entropy and dissociation time in a bottom-up holographic QCD model. For this purpose, we considered the Einstein-Maxwell-dilaton magnetised gravity model of \cite{Bohra:2019ebj}, which not only captures lattice supported inverse magnetic catalysis for the deconfinement and chiral transitions, but also anisotropic effects in the string tension. We found that in the deconfinement phase there are two string world sheet configurations, connected and disconnected, that minimize the $q\bar{q}$ free energy. These configurations  provide the usual Coulombic structure in the free energy at small separation lengths for all $T$ and $B$, whereas the free energy flattens out at large separations. This result is in accordance with lattice findings. Our analysis further suggests that similar results will persist with finite magnetic fields as well. We thoroughly analysed the temperature and magnetic field dependence of the  critical string length $\ell_{crit}$, at which a connected/disconnected phase transition appears, and found that it not only decreases with temperature but also with magnetic field for temperatures near the deconfinement temperature. Since $\ell_{crit}$ can be thought of as a rough estimate of the maximal possible separation length at which the $q\bar{q}$ pair forms a bound state, our analysis thus suggests a decrement in the size of the bound state near the deconfinement temperature as the magnetic field increases. We further found that this is true irrespective of the orientation of the magnetic field.

We then analysed the entropy of the $q\bar{q}$ pair in our model. We found not only a sharp rise and a peak in the entropy of $q\bar{q}$ pair near the deconfinement temperature but also that the entropy saturates to a constant value at interquark separations for all $B$. For $B=0$, these findings are again in qualitative agreement with known lattice QCD results. Interestingly, the magnitude of the $q\bar{q}$ entropy is found to be decreasing substantially with $B$ near the deconfinement temperature. Similarly, the saturation value of the $q\bar{q}$ entropy at large separation is found to be decreasing with $B$. These are novel predictions of our holographic model.  We further found that the strength of the entropic force increases with magnetic field near the deconfinement temperature, suggesting a stronger dissociation of heavy quarks in the presence of a magnetic field at that temperature. These results point to the inverse magnetic catalysis nature of the QCD system at the level of quarkonium bound states. In principle, the aforementioned AdS/QCD predictions could be checked against (not yet done) lattice QCD simulations.

We have also investigated the real-time dynamics of quarkonium dissociation. We calculated the dissociation time for different temperatures and magnetic fields for large interquark separations and found that the dissociation time gets smaller for higher temperatures and magnetic fields. In particular, the magnetic field enhances the dissociation rate. Again, the holographic QCD system exhibited these features for both parallel and perpendicular orientations of the magnetic field.

We end this discussion by pointing out some problems which would be interesting to analyze in the future. The important one would be to investigate how the QCD system thermalizes in the presence of a magnetic field, as the associated thermalization time scale could help us to find a more accurate picture of the quarkonium dissociation. This would again be very challenging as it would require not only a consistent construction of AdS-Vaidya type metric \cite{Balasubramanian:2011ur,Ishii:2014paa} having a background magnetic field, but also computationally it would require detailed care of spatial variational part of the world sheet equation of motion. A similar numerical algorithm could also be used to discuss the dissociation time for all interquark separation lengths, without dropping the spatial variation of the string. As such, an improved estimate of the dissociation time would be obtained from the time it takes for the connected string to change into the disconnected configuration. Another interesting problem would be to analyse the anisotropic entanglement structure of the holographic QCD system in the presence of a magnetic field, along the lines of \cite{Dudal:2016joz,Dudal:2018ztm,Mahapatra:2019uql,Jain:2020rbb}. We hope to report on these and other related topics in the near future.

\section*{Acknowledgments}
The work of S.~M.~is supported by the Department of Science and Technology, Government of India under the Grant Agreement number IFA17-PH207 (INSPIRE Faculty Award). The work of S.~S.~J.~is supported by Grant No. 09/983(0045)/2019-EMR-I from CSIR-HRDG, India. We are grateful to
O. Kaczmarek for the permission to take the Figures 1 and 2 from \cite{Kaczmarek:2005zp}.


\end{document}